\documentclass[final,5p,times,twocolumn]{elsarticle}
\usepackage{times}
\usepackage{graphicx}
\usepackage{epsfig}
\usepackage{amssymb}

\usepackage{lineno}

\usepackage{ulem}

\usepackage{natbib}

\journal{Planetary and Space Science}

\begin{document}

\begin{frontmatter}

\title{Against the biases in spins and shapes of asteroids}

\author[label1]{A. Marciniak}
\address[label1]{Astronomical Observatory Institute, Faculty of Physics, A. Mickiewicz University, 
         S{\l}oneczna 36, 60-286 Pozna{\'n}, Poland}
\ead{am@amu.edu.pl}

  \author[label2]{F. Pilcher}
  \address[label2]{4438 Organ Mesa Loop, Las Cruces, New Mexico 88011 USA}
  
  \author[label1]{D. Oszkiewicz} 
  \author[label1]{T. Santana-Ros} 
  \author[label3]{S. Urakawa}
  \address[label3]{Bisei Spaceguard Center, Japan Spaceguard Association, 1716-3, Okura, Bisei, Ibara, 
	   Okayama 714-1411, Japan}
  
  \author[label4,label5]{S.~Fauvaud}
  \address[label4]{Observato{\'i}re du Bois de Bardon, 16110 Taponnat, France}
  \address[label5]{Association T60, 14 avenue Edouard Belin, 31400 Toulouse, France}
  
  \author[label6]{P. Kankiewicz}
  \address[label6]{Astrophysics Division, Institute of Physics, Jan Kochanowski University, 
           {\'S}wi{\,e}tokrzyska 15, 25-406 Kielce, Poland }

  \author[label1]{{\L}. Tychoniec} 
  \author[label4,label5]{M.~Fauvaud} 
  \author[label1]{R. Hirsch} 
  \author[label1]{J.~Horbowicz} 
  \author[label1]{K. Kami{\'n}ski}  
  \author[label1]{I. Konstanciak} 
  \author[label7,label8]{E. Kosturkiewicz}
  \address[label7]{Mt. Suhora Observatory, Pedagogical University, Podchor{\,a}{\.z}ych 2, 30-084, Cracow, Poland} 
  \address[label8]{Astronomical Observatory of the Jagiellonian University, Orla 171, 30-244, Cracow, Poland} 
 
  \author[label1]{M.~Murawiecka} 
  \author[label1,label9]{J. Nadolny} 
  \address[label9]{Universidad de La Laguna, Dept. Astrofisica, E.38206 La Laguna, Tenerife, Spain}

  \author[label3]{K.~Nishiyama}
  \author[label3]{S. Okumura} 
  \author[label1]{M. Poli{\'n}ska} 
  \author[label5]{F.~Richard} 
  \author[label3]{T. Sakamoto} 
  \author[label1]{K. Sobkowiak} 
  \author[label7]{G.~Stachowski} 
  \author[label1]{P. Trela}

 \begin{abstract}
 
 Physical studies of asteroids depend on an availability of lightcurve data. Targets that are easy to observe and analyse 
 naturally have more data available, so their synodic periods are confirmed from multiple sources. Also, thanks to availability 
 of data from a number of apparitions, their spin and shape models can often be obtained, with a precise value of 
 sidereal period and spin axis coordinates.
 
 Almost half of bright (H$\leq$11 mag) main-belt asteroid population with known 
 lightcurve parameters have rotation periods considered long (P$\geq$12 hours) and are rarely chosen for photometric 
 observations. There is a similar selection effect against asteroids with low lightcurve amplitudes (a$_{max}\leq$0.25 mag). 
 As a result such targets, though numerous in this brightness range, are underrepresented in the sample 
 of spin and shape modelled asteroids. In the range of fainter targets such effects are stronger.
 These selection effects can influence what is now known about asteroid spin vs. size distribution, on asteroid internal 
 structure and densities and on spatial orientation of asteroid spin axes.

 To reduce both biases at the same time, we started a photometric survey of a substantial sample of 
 those bright main-belt asteroids that displayed both features: periods longer than 12 hours, and amplitudes that did not exceed 
 0.25 magnitude. First we aim at finding synodic periods of rotation, and after a few observed apparitions, obtaining spin 
 and shape models of the studied targets.

 As an initial result of our survey we found that a quarter of the studied sample (8 out of 34 targets) have rotation periods 
 different from those widely accepted. We publish here these newly found period values with the lightcurves. 
 
 The size/frequency plot might in reality look different in the long-period range. 
 Further studies of asteroid spins, shapes, and structure should take into account serious biases that are present 
 in the parameters available today.
 Photometric studies should concentrate on such difficult targets to remove the biases and to complete the sample.

\end{abstract}

\begin{keyword}
asteroids, lightcurves, selection effects
\end{keyword}

\end{frontmatter}

\section{Introduction}

 Spin and shape parameters of a large sample of main belt asteroids are an important basis
 for theories describing Solar System formation and evolution, with non-gravitational forces influencing
 the orbital and physical properties of these minor bodies. It has been recently found that asteroids 
 belonging to the Flora family have preferential prograde rotation, because retrograde rotating objects were 
 moved by the Yarkovsky effect to the $\nu_6$ resonance at the inner main belt and removed \citep{Kryszczynska_2013}.
 This fact finds its confirmation in  preferential retrograde rotation of Near Earth Asteroids\citep{La_Spina_etal_2004}.
 It has also been found that the distribution of known spin axis positions of small asteroids shows a trend for them 
 to group at extreme values of obliquities (at high angles from their orbital planes), what can be explained as the outcome 
 of spin evolution under the influence of YORP effect \citep{Hanus_etal_2013}.

 However, the available sample of well studied asteroids is burdened with substantial selection effects. 
 There exist well known strong observational biases against small, low-albedo, and distant objects due to limitations 
 of instruments that are most widely used for photometric studies.
 But there are also other strong selection effects that are connected to photometric studies. 
 \begin{table*}[h]
 \begin{scriptsize}
\begin{tabular}{|cccc|ccc|}
\hline
\hline
        & \multicolumn{3}{c}{Period of rotation} &  \multicolumn{3}{c}{Maximum amplitude}\\
\hline
        & long, P$\geq$12 h & short, P$<$12 h& undefined &  high, a$_{max}>$0.25 mag &  low, a$_{max}\leq$0.25 mag & undefined\\
        &   (modelled)      &   (modelled)   &           &   (modelled)              &      (modelled)             &          \\
\hline
Number  &    528 (101)      &    656 (236)   &   46      &   655 (233)               &   543 (104)                 &   32     \\
\hline
\hline
\end{tabular}
 \caption{Numbers of asteroids with H$\leq$11 mag, for which LCDB 
\citep{Warner_etal_2009}
 gives physical parameters.
 Only lighctcurve parameters with period quality code better than or equal to 2- have been included.
 The total number here is 1230 asteroids, median period is 10.986 h, median a$_{max}$=0.27 mag.
 Parentheses give numbers of spin and shape modelled objects within each group.}
  \label{stats}
  \end{scriptsize}
\end{table*}

\begin{figure*}[h]
\includegraphics[width=0.5\textwidth]{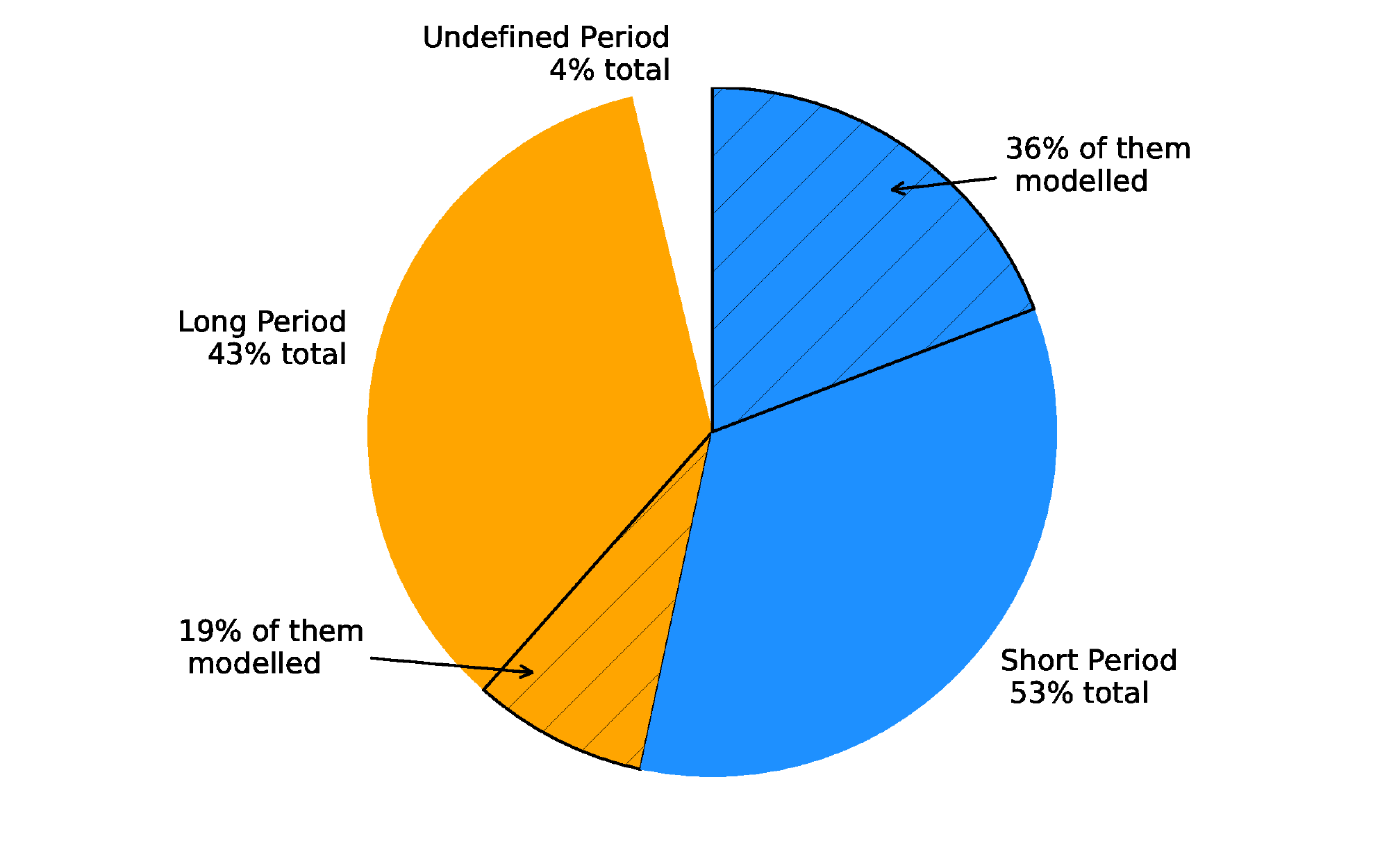}
\includegraphics[width=0.5\textwidth]{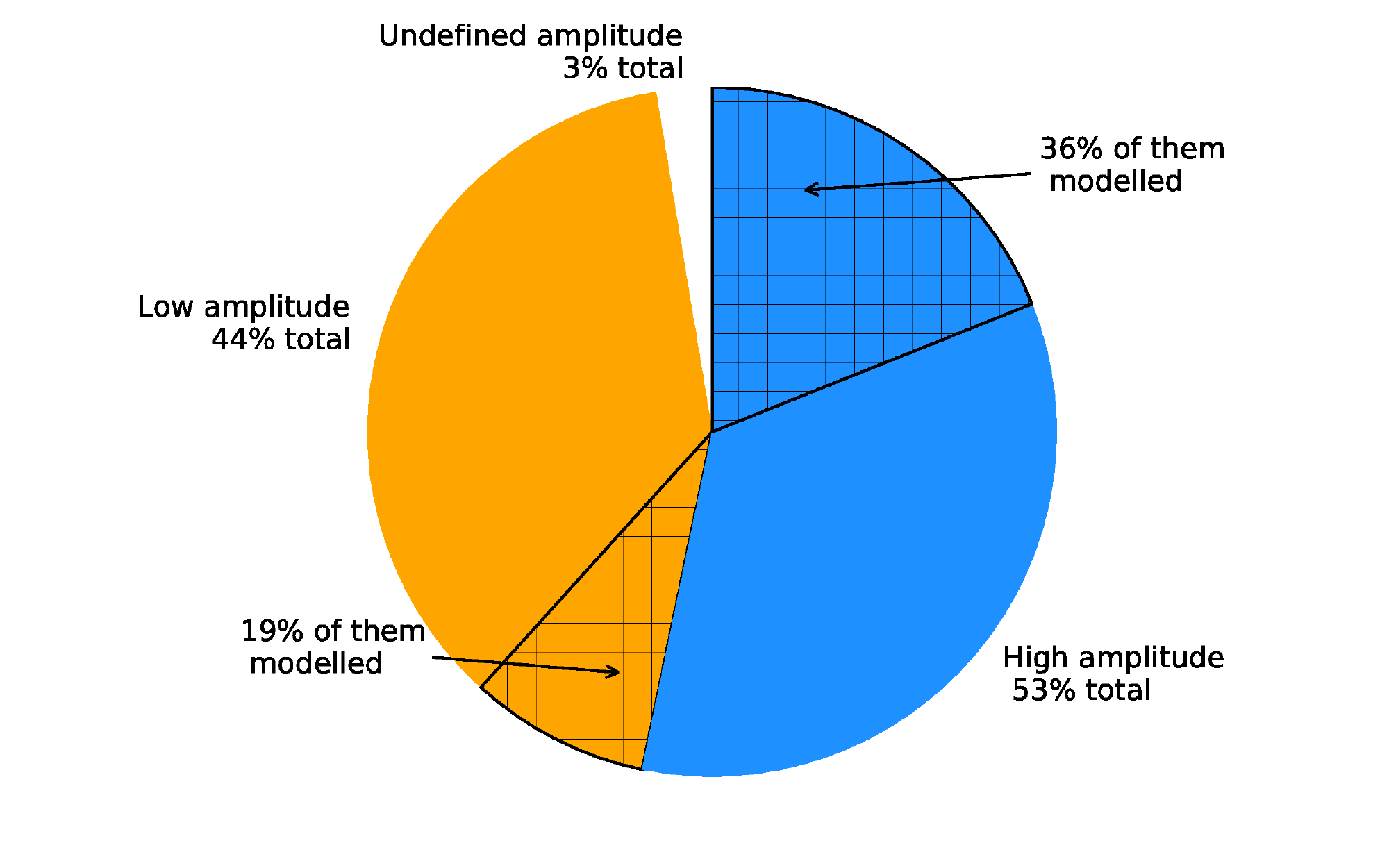}
\caption{Statistics of periods and amplitudes of bright main-belt asteroids, based on data from table \ref{stats}.}
\label{bias}
\end{figure*}

\begin{figure*}[h]
\vspace{1cm}
\includegraphics[width=0.8\textwidth]{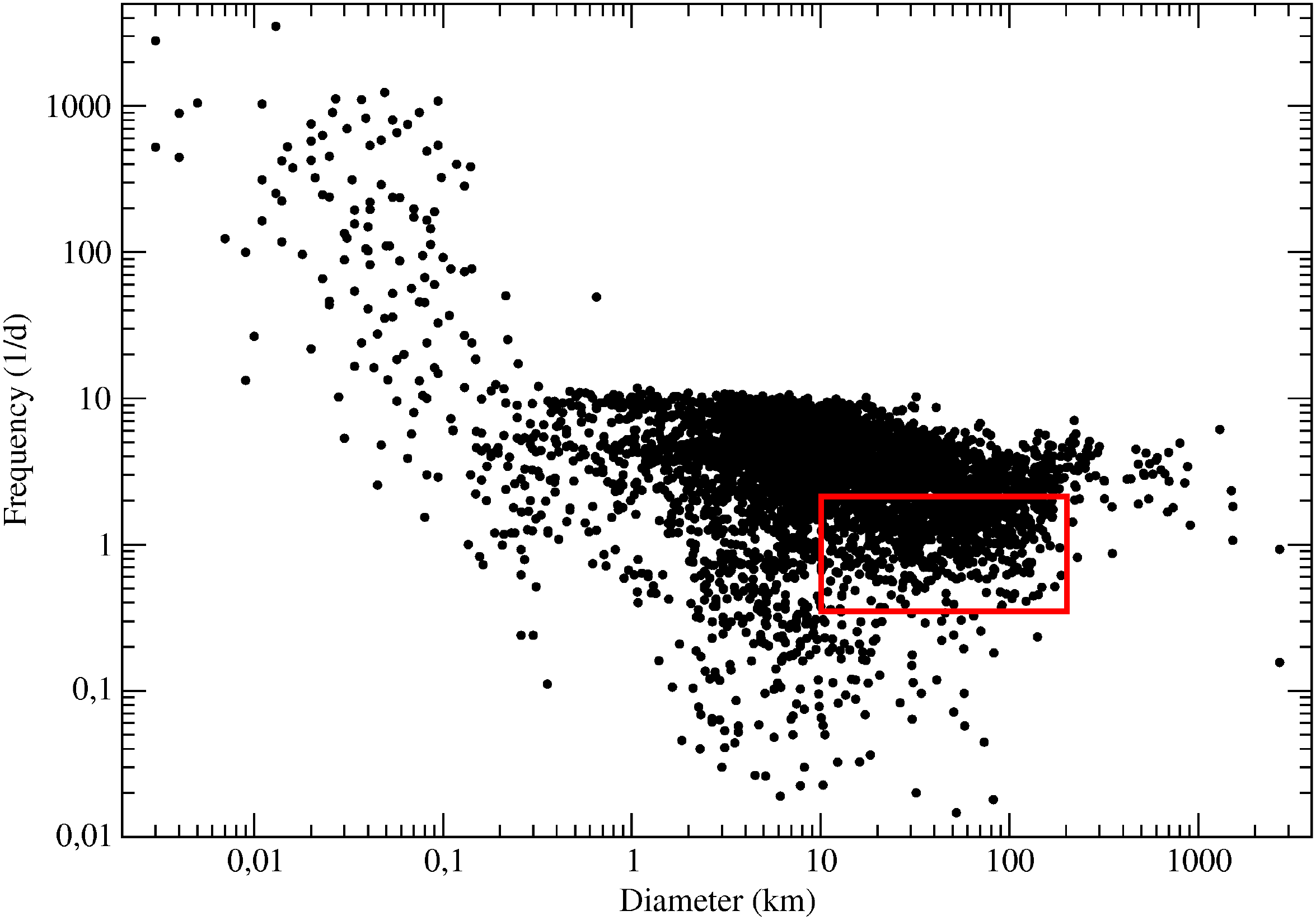}
\caption{A diagram of asteroid rotational frequencies versus sizes 
(after: \citet{Warner_etal_2009}). 
Rectangular frame highlights class of objects studied within this project.}
\label{lcdb_all_long}
\end{figure*}

 
  The arithmetic mean of periods in the size range limited by absolute magnitude H$\leq$11 mag given by 
\citet{Pravec_Harris_2000} 
  was around 9-10 hours, with a most extreme value of the mean equal to 12 hours 
  in the size range around 100 km. With time the sample of asteroids with known periods substantially grew, 
  and a large number of slow periods was found. Currently the arithmetic mean of rotation periods in this size 
  range is almost 20 hours.
  Median value of all known periods in this sample of main-belt objects with H$\leq$11 mag (where 
  the period survey is almost complete) is around 11 hours. 
  It means that half of the bright main-belt population with known 
  lightcurve parameters have rotation periods considered ''long'' and are rarely chosen as targets for photometric observations, 
  even though they are easy targets for small telescopes. As a result only 20\% of this group has been spin and shape modelled.
  Along with the bias against asteroids with long periods of rotation (here: P$\geq$12h), there is another one, against those 
  with low lightcurve amplitudes (here: $a_{max}\leq$0.25 mag).
  Within the set of bright asteroids those with small amplitudes comprise almost half of the whole studied population,
 while spin and shape models have been determined for around 20\% of this sub-population 
 (source: LCDB; \citet{Warner_etal_2009}, 
see table \ref{stats}). 
 On the other hand two remaining populations (short-period, and high-amplitude objects) have been modelled 
 in 36\% each (see fig. \ref{bias}). 
 By chance, at such conditions statistics in both groups, divided by period and by maximum amplitude, are very similar 
 and both diagrams look almost identical to each other.
 For fainter targets these inequalities are much stronger, and the data pool of objects with known periods is highly 
 incomplete. Moreover, the widely known plot showing distribution of rotational frequency vs. asteroid sizes might in reality 
 look slightly different in the lower-right part (fig. \ref{lcdb_all_long}).

\begin{figure*}[h]
\includegraphics[width=0.7\textwidth]{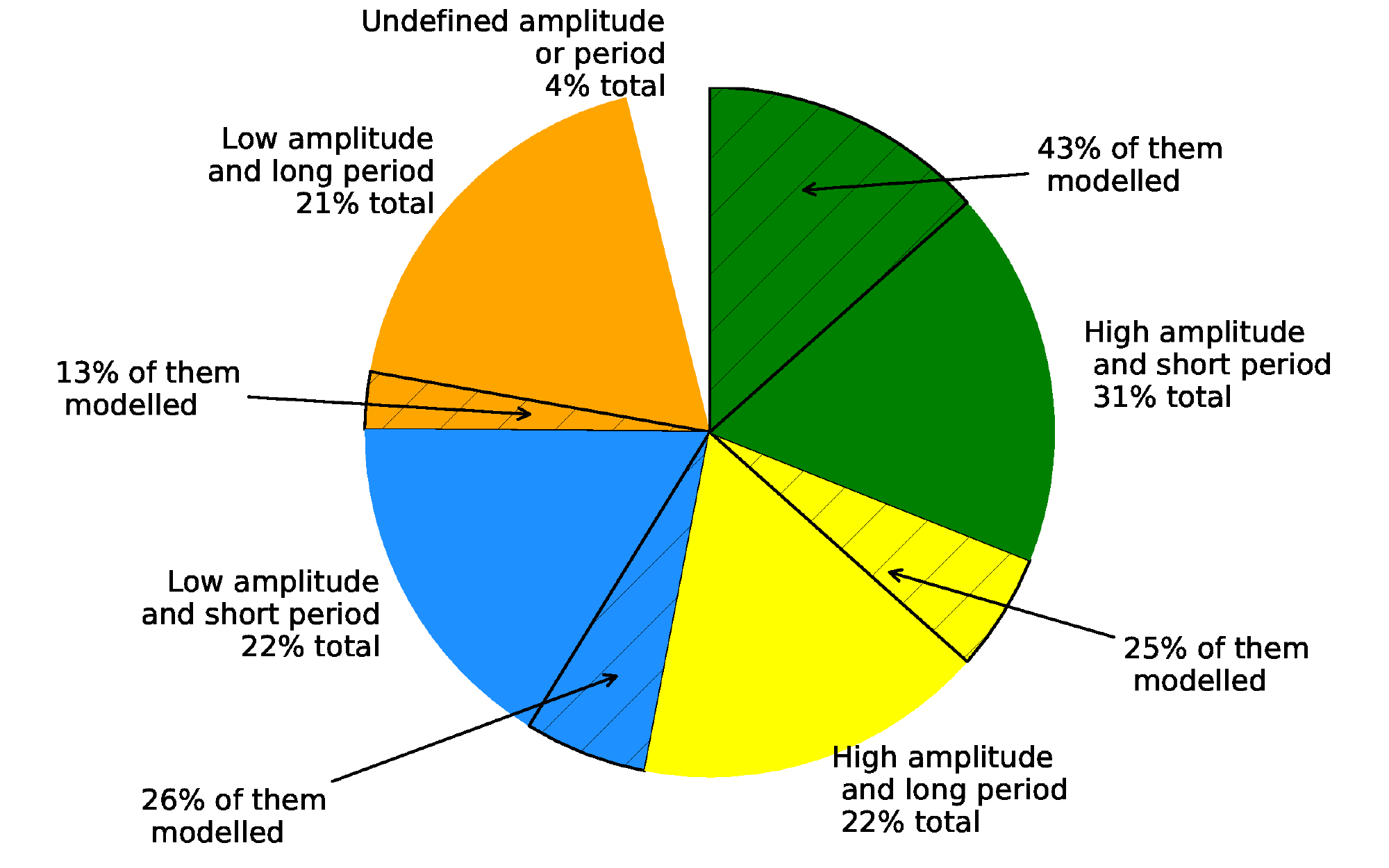}
\caption{Statistics of periods and amplitudes of bright main-belt asteroids based on two parameters simultaneously, 
based on data from table \ref{stats2}}
\label{bias3}
\end{figure*}

  In the sample of main-belters with H$\leq$11 mag and with available lightcurves, 
  only 13\% of long-period objects with low amplitudes have been 
  spin and shape modelled, while 43\% of the comparably numerous group with opposite features 
  (periods shorter than 12 hours and maximum amplitudes higher than 0.25 mag) have been modelled 
  (see table \ref{stats2} and fig. \ref{bias3}). 
  It means that almost half of the easily observable population is well studied, 
  while objects more difficult to properly observe 
  and analyse are well studied in a much smaller fraction of their population.
 \begin{table*}[h]
 \begin{scriptsize}
 \begin{tabular}{|cccccc|}
 \hline
 \hline
        & High amplitude & High amplitude & Low amplitude   & Low amplitude & Undefined\\
        & \& short period & \& long period & \& short period & \& long period& amplitude or period\\
 \hline
 Number &    382 (165)   &    270 (68)   &   272 (71)      &    258 (33)   &   48     \\
 \hline
 \hline
 \end{tabular}
 \caption{Same as table \ref{stats} but for two parameters simultaneously. Parentheses give numbers of spin and shape modelled objects 
  within each group.}
  \label{stats2}
 \end{scriptsize}
 \end{table*}

  The observational bias against asteroids with small amplitude of brightness variations can influence our knowledge of 
  the asteroid spin axis distribution in space, on their shapes and densities. Since weakly elongated objects 
  always display small lightcurve amplitudes, their periods are hard to determine, and their spin and shape models are 
  very hard to obtain. They become even more difficult for modelling when their spin axes are close to their orbital 
  planes, causing any brightness variations disappear when asteroid is viewed nearly pole-on 
\citep{Marciniak_Michalowski_2010}. 
  Moreover, data for low-amplitude objects can be lost in the 0.1 - 0.2 magnitude noise of the sparse data 
  coming from large, astrometric surveys \citep{Hanus_etal_2011}, 
 making them hard to be modelled using such data alone. 
  Photometric data for unique spin and shape modelling of low-amplitude asteroids should be at least an order of magnitude 
  better, with the noise not exceeding 0.01 mag. 

  Small, kilometer sized asteroids with long periods of rotation can be those that have been spun down by YORP torques 
  \citep{Rubincam_2000} 
 or can be in tumbling rotational states, showing double periodicity. 
  Thus they are important targets for studies of damping timescales and evolution under thermal effects. 
  Additionally, studies in the small amplitude range can result in possible discoveries of binary systems, 
  because in majority of binaries the components are weakly elongated in shape 
\citep{Pravec_Harris_2007}. 

  Summing up, it turns out that at least half of the whole well observable main belt population is studied insufficiently, 
  and the selection effects tend to increase with time. Previously found groupings and depopulations within  
  large number of asteroid spin axis longitudes could have been caused by several overlapping selection effects 
\citep{Bowell_etal_2014}. 

Also, it has been shown that even precise data from the Gaia satellite will introduce biases 
  in derived asteroid spin and shape models depending on their true pole latitude and shape elongation 
\citep{Santana-Ros_etal_2015}. 

\section{Observing campaign}

  Motivated by these facts we started a large, long-term observing campaign of a substantial sample of long-period asteroids
  from those objects (highlighted with rectangular frame in fig. \ref{lcdb_all_long}) that additionally displayed small 
  amplitudes of brightness variations, to reduce both selection effects at the same time. 
  Our aim is to obtain spin and shape 
  models of this class of ''difficult'' objects in order to reduce the increasing biases favouring quickly
  rotating, elongated asteroids, with extreme values of spin axis obliquity to be prevailingly 
  observed and modelled today. It is caused by the fact that the latter objects are easiest to observe and analyse 
  - their short periods allows for a full lightcurve coverage in one or two nights, and large amplitudes makes 
  their brightness variations always stand out of the noise even in imperfect observing conditions.
  On the other hand, obtaining full phase coverage for objects with long periods requires much more telescope time 
  and, if their amplitude is small, data of higher quality. The periods of the order of several dozens of hours 
  have to be determined from fragments covering only a small fraction of a full rotation. If the lightcurves 
  are symmetric - i.e. both maxima look similar - making a composite lightcurve can often lead to finding an 
  alias period like 2/3 or 3/2 of the true period. This might be a potential source of another bias: asteroids with 
  long periods, even if well observed, can have their periods determined incorrectly. This effect can be even stronger 
  for objects displaying both features  - long period and low amplitude - at the same time.

   We restricted our statistical studies of to the sample of main-belt asteroid brighter than 11 absolute magnitude, 
 because in this group the survey of lightcurve parameters is almost complete. This set roughly coincides with 
 the first thousand of numbered asteroids. Also, it has been shown that all main-belt objects have been discovered 
 down to the absolute magnitude 11.25 
\citep{Jedicke_Metcalfe_1998}. 
 A great majority of these asteroids are objects with bright apparent magnitude (not fainter than 14 mag), 
 well within reach of small amateur telescopes. Still, many of them are insufficiently studied, not even allowing for 
 a unique period determination. There is a clear correlation between the number of apparitions in which an asteroid has been 
 observed and its period. When the period is known to be longer than 10 - 12 hours it is avoided by most observers. 
 It is understandable, the telescope time is often very limited, and such objects often cannot be observed over full rotation 
 from one site for example due to certain commensurabilities with Earth day. 
  
 As for the bias against small amplitudes, the limitations here are of a different nature. Quite often lightcurve 
 amplitudes of such objects are at the level of a few hundreths of a magnitude. It requires good instrumental and weather 
 conditions, so that the data are not burdened with noise greater than a few thousandths of a magnitude. Since small telescopes 
 and very short exposure times are most widely used, this level of precision cannot always be reached. Sometimes the data can 
 be binned to decrease the noise - an efficient strategy for objects with long periods. It is also possible that 
 seemingly flat lighcturves were considered uninteresting and left unpublished, while simple rescaling of the plot 
 (stretching the magnitude scale and compressing the time scale) would 
 show them to be clear fragments of a long-periodic lightcurve, where adding a few more such fragments would suffice 
 for completion. We encourage observers to submit such data to ALCDEF repository\footnote{ \tt http://www.minorplanet.info/alcdef.html}, 
 even if incomplete, they can be very useful for asteroid modelling.

 Our target selection procedure was as follows: 
 all the long-period (P$\geq$12h) and low-amplitude (a$_{max}\leq$0.25 mag, or 0.35 mag in single cases)
  asteroids without spin and shape models were included from 
 the list of main belt asteroids with absolute magnitude brighter or equal to 11 mag, with the exception of objects 
 belonging to certain groups and families that are studied by other researchers to avoid duplication, and single cases 
 of very long period objects (P$>$50h), for practical reasons. Asteroids with periods longer than 50 hours present only 3\% 
 of the studied population. While they deserve to be studied for many reasons like a possibility of tumbling rotational state 
 or a binarity, a separate large observing campaign would be needed for them. 
 Still, since they comprise only a small part of the studied population, we don't introduce a large bias by omitting them.
 On the other hand we included objects with unknown or uncertain period, because that most probably means ``long''. 
 The bias that we conserve here is the one against faint targets, because of the equipment limitations. 
 But with an access to larger telescopes, we are going to address this bias as well, extending our 
 target list to fainter objects, where the selection effects studied here are even more profound.
 As of now our target selection resulted in the list of 120 bright main-belt asteroids (almost half of the population 
 of long-period and low-amplitude objects within this brightness range) and up till December 2014 we gathered data 
 for 45 of them (see the Results section).

For the needs of this project we set up an international network of observatories, with the local observing station 
at the Borowiec Station of Pozna{\'n} Astronomical Observatory Institute (Poland) being the main site.
Other observatories are: Observatori Astron\`o{}mic del Montsec - OAdM - in Catalonia (Spain); 
Organ Mesa Observatory in New Mexico (USA); 
Winer Observatory in Arizona (USA); 
Bisei Spaceguard Center (Japan)\footnote{Bisei Spaceguard Center is administrated by the Japan Space Forum}. 
We also get an occasional support from
Jan Kochanowski University in Kielce and Mnt. Suhora observatories (Poland), and Observatoire du Pic du Midi (France).
Their locations and telescope parameters are summarised in table \ref{telescopes}.
  Such network allows for effective coordination of the observing campaign, for example for asteroids with periods close
 to 12 or 16 hours, or their integer multiples. The crop of the observing campaign conducted over last two years 
 consists of 1670 hours of data.
 \begin{table*}[h]
   \begin{scriptsize}
 \begin{tabular}{|ccccc|}
 \hline
 \hline
  Observatory name & abbreviated name & IAU code & location & telescope diameter\\
 \hline
  Borowiec Observatory (Poland) & Bor. & 187 & 52 N, 17 E & 0.4m \\

  Montsec Observatory (Catalonia, Spain) & OAdM & C65 & 42 N, 01 E & 0.8m\\

  Organ Mesa Observatory (NM, USA) & Organ M. & G50 & 32 N, 107 W & 0.35m\\

  Winer Observatory (AZ, USA) & Winer & 648 & 32 N, 111 W & 0.70m\\

  Bisei Spaceguard Center (Okayama, Japan) & Bisei & 300 & 35 N, 134 E & 0.5m and 1m \\

  JKU Astronomical Observatory (Kielce, Poland) & Kie. & B02 &  51 N, 21 E & 0.35m\\

  Mt. Suhora Astronomical Observatory (Poland) & Suhora & - & 50 N, 20 E & 0.25m and 0.60m\\

  Pic du Midi Observatory (France) & Pic & 586 & 43 N, 0 E & 0.6m\\
 \hline
 \hline
 \end{tabular}
 \caption{Observatories participating in this project, with their abbreviated names, IAU codes, coordinates, and telescope diameters}
  \label{telescopes}
 \end{scriptsize}
 \end{table*}

 Because of the small amplitudes of lighcturves, good quality photometry is neccessary, with the noise at the level 
 of a few tenths of a percent. Such quality data are being regularly obtained in the Borowiec Station of Poznan Astronomical 
 Observatory Institute. During more than fifteen years of station operation optimal observing and reduction procedures 
 have been established and substantial experience has been gained both in asteroid photometry and modelling 
 (see \citep{Michalowski_etal_2004} 
for the description of the station and reduction procedures, and e.g. 
\citet{Marciniak_etal_2012} 
 for lightcurves and models).

 We perform photometric observations of asteroids using Red and Clear filters, with standard bias, dark frame, and flatfield 
 corrections. At the next stage the aperture photometry is applied with a high level of automatisation, but with a careful 
 control of a comparison star colours and possible intrinsic variablity; a cross check between at least three different 
 comparison stars is performed to exclude it.
 Also, any field star passages and possible interfering artifacts, like satellite trails and seeing changes 
 are checked, together with other possible instrumental problems. 
 A resulting fragment of a relative lightcurve is binned when necessary, so that the number of datapoints per period 
 would not exceed 200. 

  It might seem that in case of long-period objects performing an absolute photometric measurements is the right choice. 
  However we rely on relative rather than absolute photometric measurements, for the following reasons.
  First of all the observing conditions in our local observing station are rarely photometric, making 
  measurements of absolute magnitudes burdened with serious errors.
  But the main reason is that we are in the small amplitude range, with amplitudes of some objects 
  being below 0.1 mag. Having the uncertainties of absolute magnitude measurements normally at the level of 
  a few hundreths of a magnitude would lead to a situation where the measurement uncertainties are of the same order as 
  the determined values (amplitudes). In the end an arbitrary shifting in the magnitude scale would be neccessary, 
  meaning that the absolutisation was done in vain. 
 
 Our observing campaign relies on optimised observing strategy and careful planning of possible phase coverage from 
 multiple stations. 
 We aim at obtaining at least one long lightcurve fragment and completing the whole rotation with shorter 
 fragments, or of similar in length, when the observing curcumstances allow. 
 The data are reduced and composite lightcurves are made shortly after an observing session, 
 so that new observations can be planned when an asteroid can still be reached. When a new period is found the campaign 
 is intensified, focusing on this object to confirm the new value and exclude any possible alias periods.
 When the period is well known we are also trying to register the phase angle effects, like e.g. growing amplitude, 
 what requires observing at least 1/4 of the full rotation in a single session. However with periods exceeding 30 hours, 
 that is often impossible. For a precise period determinations we rely on the repeatable fragments of the lightcurves 
 and realistic constraints put on the lightcurves with higher harmonics 
\citep{Harris_etal_2014}. 

  Our group has access to the 80-cm robotic telescope in OAdM Observatory in the Montsec range on the south-west 
  side of the Pyrenees. 
  To increase the return of data gathering on such an instrument we are utilising the 
  100 hours granted each year with the observing strategy of obtaining single images of each object at half hour intervals. 
  It facilitates observing a greater number of long-period objects in a much shorter time compared to traditional continuous 
  photometry, at the same time not altering the details of the resulting lightcurves, as we repeat such observations until 
  we gather at least 100 points per period. The acquired know-how served in similar observations on another robotic telescope: 
  the 0.7m GATS-PST2 telescope built in our institute in Poland and recently moved to Winer Observatory in Arizona.

\section{Results}

  At this stage none of the asteroids observed in our programme has enough data for a unique 
  spin and shape model. However, as an intermediate goal, we determined synodic periods for all the objects observed. 
  To our own surprise we found periods differing from those previously accepted for as many as 8 
  out of 34 well observed asteroids (those for which the amount of data allows for a unique period determination; 
  see table \ref{Results}).
  The quality codes for their periods in LCDB summary line were ''2'' (period certain within 30\%, or to an integer ratio), 
  but also ''3'' which means a secure period determination.  
  From the remaining number of 11 studied objects at least 2 also show values of periods different from those in the LCDB, 
  but the amount of data we gathered for them does not allow for making firm conclusions yet.
  We decided to publish here those newly found periods that we consider secure, together with composite lightcurves for these 
  8 asteroids. They are a significant fraction of the sample (24\%) studied by us, confirming that this class of objects 
  is burdened with serious bias in basic physical parameters in addition to the observing biases that prevent most of them 
  from being studied according to the number in their population.
  Studied here objects make up around 1/5 of the whole population of bright main-belt asteroids.
  Since 1/4 of the sample that we studied showed to have 
  different periods to those widely accepted, we can assume that at least 1/20 of the main-belt 
  population probably still have their periods determined incorrectly (this would mean a few dozen objects from the first 
  thousand of numbered asteroids).
  It applies even more to fainter objects, where the uncertainties are greater and where many do not have any 
  period determinations.
\begin{figure*}[h]
\vspace{0.9cm}
\includegraphics[width=0.7\textwidth]{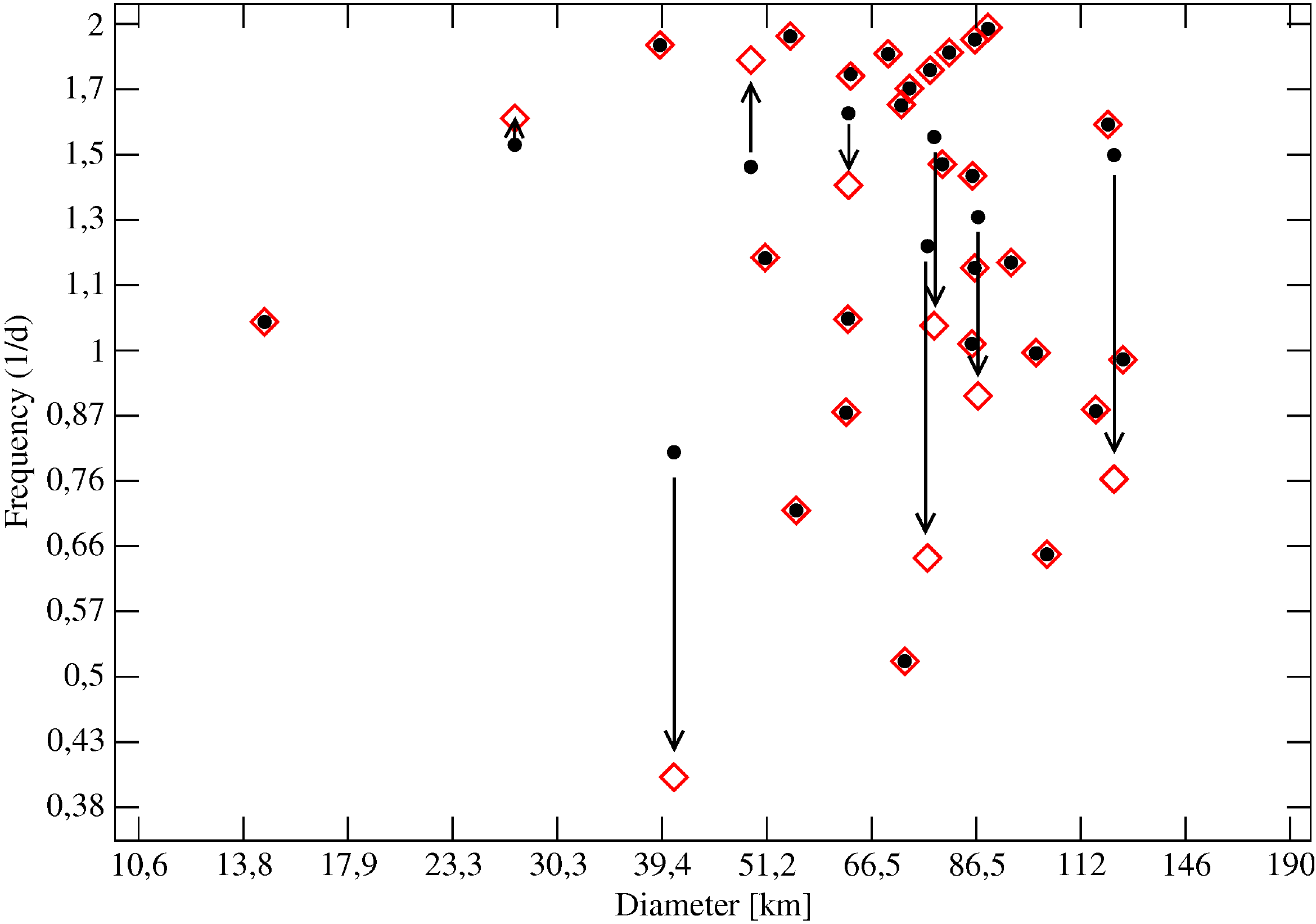}
\caption{Rotation rate (frequency) versus diameter of the objects studied. 
Filled circles denote data based on LCDB, diamonds denote frequencies found in this work, while diameters remain the same.
The scale coincides with the rectangular frame from fig. \ref{lcdb_all_long}.}
\label{D_f}
\end{figure*}

\begin{table*}[h!]
\begin{scriptsize}
\begin{tabular}{|l|c|c|c|c|l|}
\hline
&& amplitude (LCDB && U &\\
 asteroid name  & diameter (LCDB)  & and \textit{this work})& Period (LCDB)& code & Period (this work) \\
   & \hspace{0.5cm}[km]& \hspace{0.5cm}[mag]     &\hspace{0.5cm}[h]& & \hspace{0.5cm} [h]  \\
\hline
\bf{(70) Panopaea}& 122.17 & 0.06 - \textit{0.14}& \bf{15.797} & 3 & {\bf 31.619} $\pm$ 0.007\\
(159) Aemilia  & 124.97  & 0.17 - \textit{0.26} & 24.476 & 3- & 24.485 $\pm$ 0.002\\
(172) Baucis   &  62.43 & \textit{0.21} - 0.35 & 27.417  & 3 & 27.402 $\pm$ 0.005\\
(195) Eurykleia&  85.71 & 0.10 - 0.24          & 16.521  & 3 & 16.518 $\pm$ 0.002\\
(202) Chryseis &  85.58 & 0.04 - \textit{0.28} & 23.670  & 3 & 23.668 $\pm$ 0.002\\
\bf{(219) Thusnelda}& 40.56 & 0.19 - \textit{0.24} & \bf{29.842} & 3 & {\bf 59.74} $\pm$ 0.02\\
\bf{(227) Philosophia}&86.90 & 0.06 - 0.20      &\bf{18.048}  & 2 & {\bf 26.46} $\pm$ 0.01\\
(236) Honoria  &  86.20 & 0.05 - \textit{0.27} & 12.333  & 3 & 12.338 $\pm$ 0.002\\
(301) Bavaria  &  54.27 & 0.28 - \textit{0.31} & 12.253  & 3 & 12.243 $\pm$ 0.002\\
\bf{(305) Gordonia} &  49.17 & 0.16 - \textit{0.23} &\bf{16.2} & 2 & {\bf 12.893} $\pm$ 0.002\\
\bf{(329) Svea}     &  77.83 & 0.09 - 0.26      & \bf{15.201}  & 3 & {\bf 22.778} $\pm$ 0.006\\
(335) Roberta  &  89.10 & 0.05 - \textit{0.19} & 12.054  & 3 & 12.027 $\pm$ 0.003\\
(380) Fiducia  &  73.19 & 0.04 - 0.32          & 13.69   & 3 & 13.70 $\pm$ 0.02\\
(387) Aquitania& 100.51 & \textit{0.18} - 0.25 & 24.144  & 3 & 24.13 $\pm$ 0.01\\
(395) Delia    &  50.98 & \textit{0.16} - 0.25 & 19.71   & 2 & 19.680 $\pm$ 0.005\\ 
\bf{(439) Ohio}&  76.57 & \textit{0.23} - 0.24 &\bf{19.2} & 2, A & {\bf 37.46} $\pm$ 0.01\\ 
(476) Hedwig   & 116.76 & 0.13 - \textit{0.17} & 27.33   & 3 & 27.246 $\pm$ 0.005\\
(478) Tergeste &  79.46 & \textit{0.15 - 0.30} & 16.104  & 2+ & 16.105 $\pm$ 0.001\\
(483) Seppina  &  69.37 & 0.14 - 0.29      & 12.727  & 3 & 12.719 $\pm$ 0.002\\
(487) Venetia  &  63.15 & 0.03 - 0.30      & 13.28   & 2 &  13.342 $\pm$ 0.002\\
(501) Urhixidur&  77.06 & \textit{0.12} - 0.14 & 13.1743 & 3 & 13.175 $\pm$ 0.002\\
(524) Fidelio  &  71.73 & 0.18 - 0.22      & 14.198  & 3 & 14.177 $\pm$ 0.005\\
(538) Friederike& 72.34 & 0.03 - 0.25      & 46.728  & 3 & 46.7  $\pm$ 0.3\\
(618) Elfriede & 120.37 & 0.12 - 0.20      & 14.801  & 2 & 14.800 $\pm$ 0.005\\
(653) Berenike &  39.18 & 0.03 - 0.11      & 12.4886 & 3 & 12.481 $\pm$ 0.006 \\
\bf{(666) Desdemona}& 27.22 & 0.11 - \textit{0.22} & \bf{15.45} & 2 & {\bf 14.607} $\pm$ 0.004\\
(667) Denise   &  80.85 & 0.24 - \textit{0.25} & 12.687  & 3 & 12.686 $\pm$ 0.003\\
(672) Astarte  &  14.54 & 0.10 - \textit{0.17} & 22.572  & 3 & 22.588 $\pm$ 0.005\\
(780) Armenia  &  94.40 & 0.10 - 0.18      & 19.891  & 3 & 19.89 $\pm$ 0.01\\
(788) Hohensteina&103.29& 0.12 - 0.18      & 37.176  & 3 & 37.13 $\pm$ 0.05\\
\bf{(806) Gyldenia} &  62.82 & 0.10 - 0.27 & \bf{14.45} & 2 & {\bf 16.852} $\pm$ 0.006\\
(907) Rhoda    &  62.73 & \textit{0.08} - 0.16 & 22.44  & 3-& 22.45 $\pm$ 0.01\\
(980) Anacostia&  86.19 & \textit{0.05} - \textit{0.18} & 20.117 & 3 & 20.113 $\pm$ 0.004\\
(1062) Ljuba   &  55.10 & \textit{0.11} - 0.17 &  33.8  & 3 & 33.79 $\pm$ 0.02\\
\hline
\end{tabular}
\caption{Asteroid synodic periods and amplitudes found within this project
compared to literature data gathered previously in LCDB 
\citep{Warner_etal_2009} 
as for November 2012}
\label{Results}
\end{scriptsize}
\end{table*}

In the figure \ref{D_f} we show the rotation frequencies versus diameters of the studied asteroids. One set comes 
from LCDB \citep{Warner_etal_2009} 
and is marked with filled circles, and the other set presents our results 
(new periods with the same diameters) marked with diamonds. Both sets mostly coincide, but for 8 objects we see 
substantial shift in frequency scale, mostly downwards, to lower frequencies (longer periods). 
The scale is set to cover the rectangular area from figure \ref{lcdb_all_long}.

  Figures \ref{Panopaea} - \ref{Gyldenia} show the composite lightcurves of those asteroids for which we found 
  new values of the period. These composites have been created using the procedure described by 
\citet{Magnusson_Lagerkvist_1990}. 
 The individual lightcurves have been composited with the best fitting 
synodical period written in the graphs. Points from different nights are marked with different symbols. The vertical 
position of each individual lightcurve is obtained to minimize the dispersion of data points relative to their neighbours. 
The abscissae are the rotational phases with the zero points corrected for light--time. 
 After the best value for the period was found, the composite lightcurves were tried with various values of the period close to 
 the best one (in terms of $\chi^2$). Those giving almost equally good fits of the fragments to each other were accepted, 
 until notably worse fit was found. This way the range of good periods was established and half of its value 
 was accepted as the period uncertainty.

 In addition to best fit-lightcurves we present period spectra for each of these targets 
 (see figures \ref{70periods} - \ref{806periods} in the Appendix). In all cases period found here 
 gives substantially lower RMS value than previously accepted one. 
 However in one case the period accepted here cannot be distinguished from its double. 
 See the discusion for (227) Philosophia in section \ref{(227)Philosophia} for the best period selection criteria.

  The following subsections describe in detail previous works and our findings concerning specific objects.
  Lightcurves of the remaining objects of the studied sample will be published when enough data are gathered 
  to facilitate obtaining their spin and shape models. However the synodic periods that we found for them 
  agree with the values accepted in LCDB within 1\% or less (see table \ref{Results}). 
  
 Table \ref{Results} presents our main results. After asteroid numbers and names, it gives their diameters cited after LCDB. 
 The third column contains amplitude range (a$_{min}$ and a$_{max}$) from LCDB updated with the values found within this survey, 
 marked with italics. The last two columns present periods of rotation: the one from the LCDB summary line, and 
 the period found in this study with uncertainty values. We mark with boldface those targets for which periods found here differ 
 substantially from the previously accepted values.
 In the table we cite the period values as they were in the summary line of the November 2012 public 
 release of the LCDB, when we were starting our project. 
 In the meantime, three values of the LCDB periods have changed. They are cited in respective subsections.


\subsection{(70) Panopaea}
  
 Previous observations of Panopaea were conducted by 
\citet{Schroll_etal_1983}, 
and 
\citet{Harris_Young_1989} 
 in the year 1980. Both resulted in slightly asymmetric, bimodal lighcturves of 
 0.12-0.11 mag amplitude that were composited on a basis of 15.87 and 15.797 hours period, respectively. 
 Next, Panopaea was observed by \citet{Denchev_etal_1998} 
 who only recorded an amplitude greater than 0.1 mag, 
 and by \citet{Behrend_etal_2014} 
 in 2006, when it displayed a lightcurve where one of the maxima in the 15.79-hour 
 composite was almost gone, and an amplitude was as small as 0.06 mag. 
\begin{figure}[h]
\vspace{0.8cm}
\includegraphics[width=0.49\textwidth]{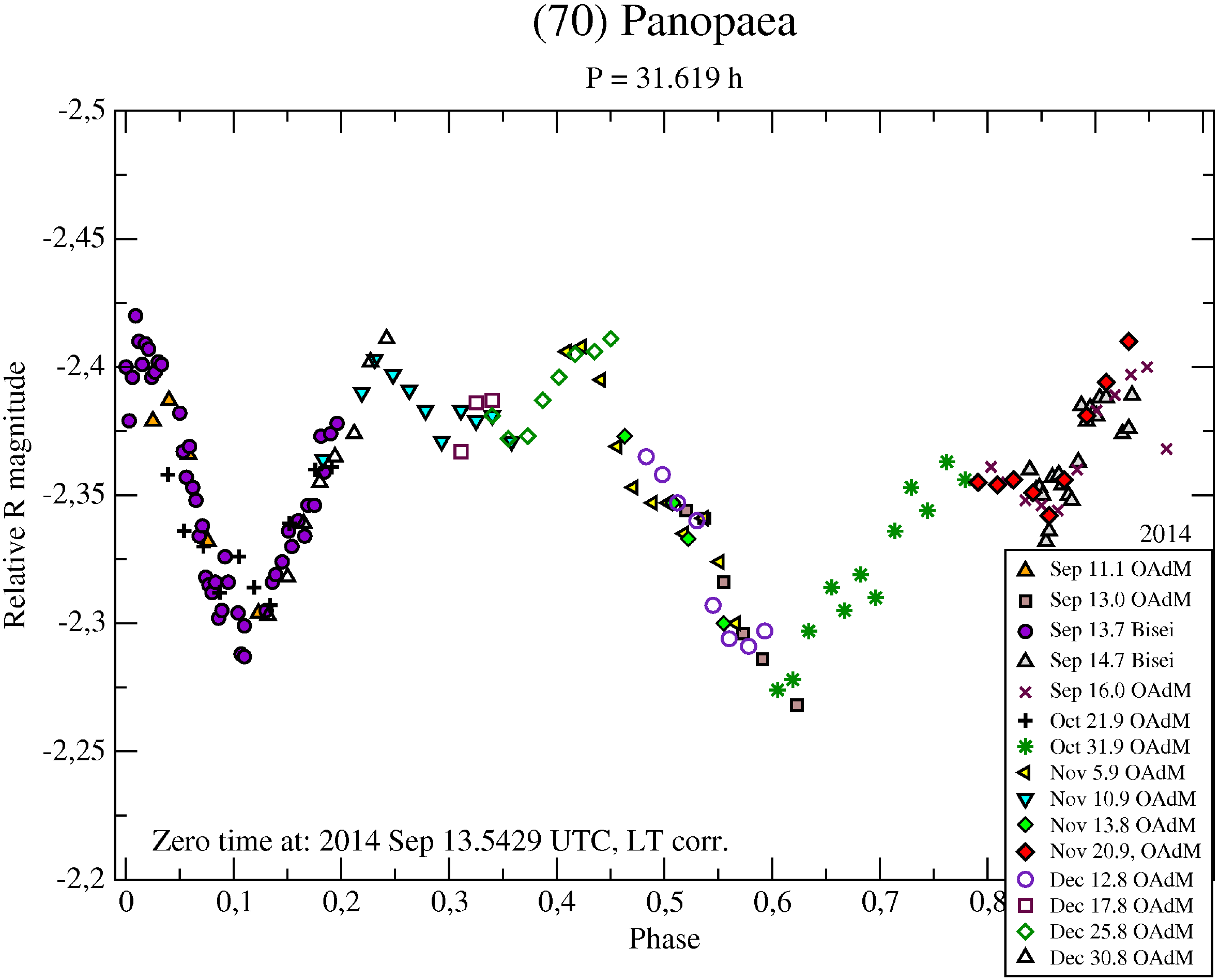}
\includegraphics[width=0.46\textwidth]{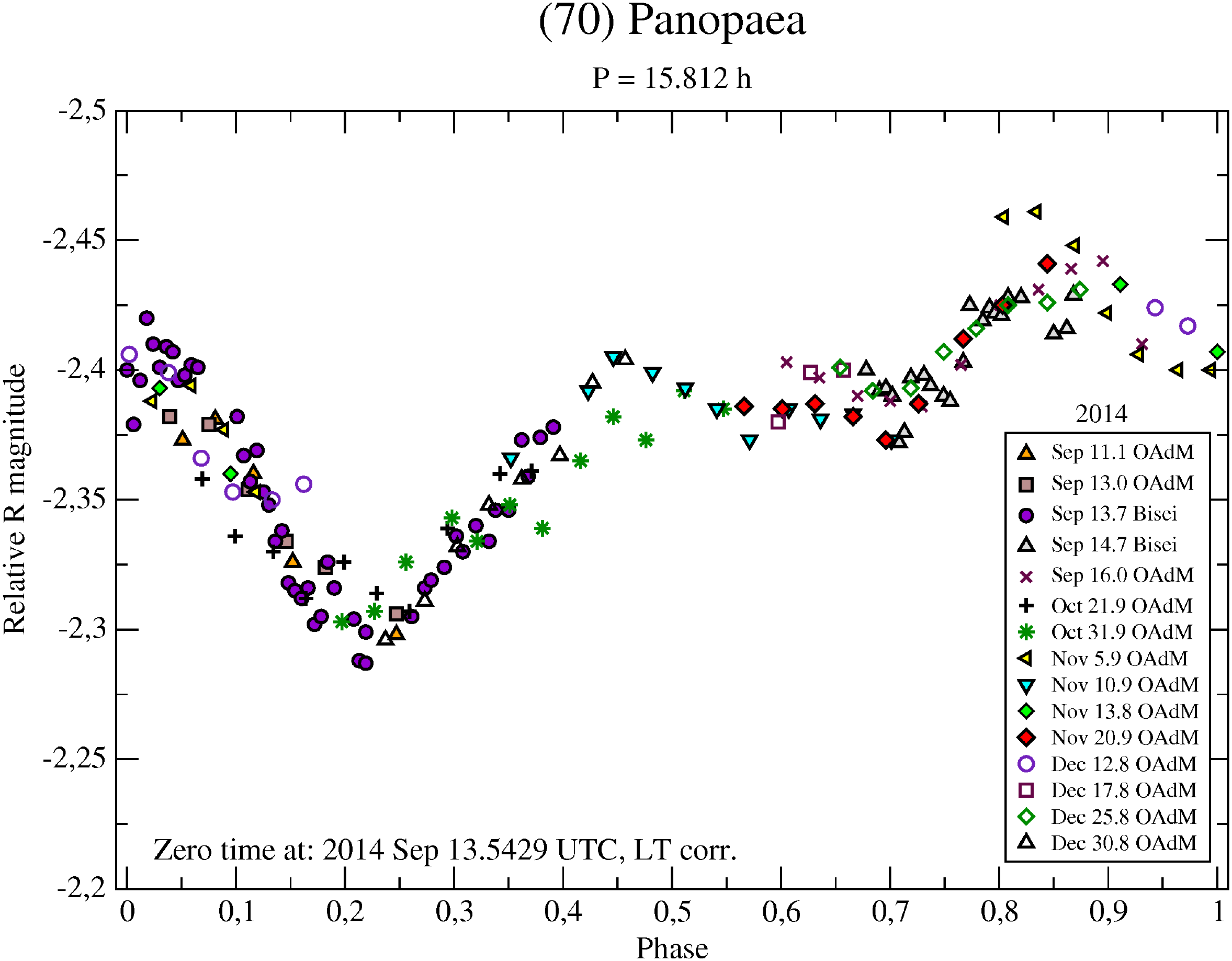}
\caption{Composite lightcurve of (70) Panopaea in the year 2014, with the best-fit period on the left
 and half period on the right.}
\label{Panopaea}
\end{figure}

 We observed (70) Panopaea from early September till late December 2014. We noticed that 15.812 hours composite 
  lightcurve (best fitting in this period range)
 gave noticeable worse fit than its double: 31.619 hours (see fig. \ref{Panopaea}). 
 For example 
 the fragments from 13.7 September (filled circles), 13.0 September (filled squares), and 31.9 October (stars) 
 have different slopes, although obtained close in time. These fragments suggest the presence of two minima:
 one narrow, and the other wider. 
 We are aware that the phase angle effects might slightly change the character of the lightcurve, but we 
 consider the apparent misfits too big for those effects acting over a few weeks. We conclude that the period 
 of this asteroid is 31.619 hours rather than 15.812, but it needs further confirmation in the next apparitions. 
 The light variations amplitude in the year 2014 was 0.14 mag ($\pm$ 0.02 mag), a small value, 
 but the biggest observed for Panopaea so far.

\subsection{(219) Thusnelda}

 This bright, main-belt asteroid was observed in the year 1981, by 
\citet{Lagerkvist_Kamel_1982}, 
and by 
\citet{Harris_etal_1992}. 
No other lighcurve has been obtained since then. 
 Both groups determined a rotation period around 29.8 hours (29.76 h and 29.842 h respectively) from around 7  
 lightcurve fragments. The amplitude was at the level of 0.20 - 0.19 mag and the composite lightcurves 
 showed one maximum with a ''shelf'' before it, which could be considered a second maximum.
\begin{figure}[h]
\vspace{0.8cm}
\includegraphics[width=0.48\textwidth]{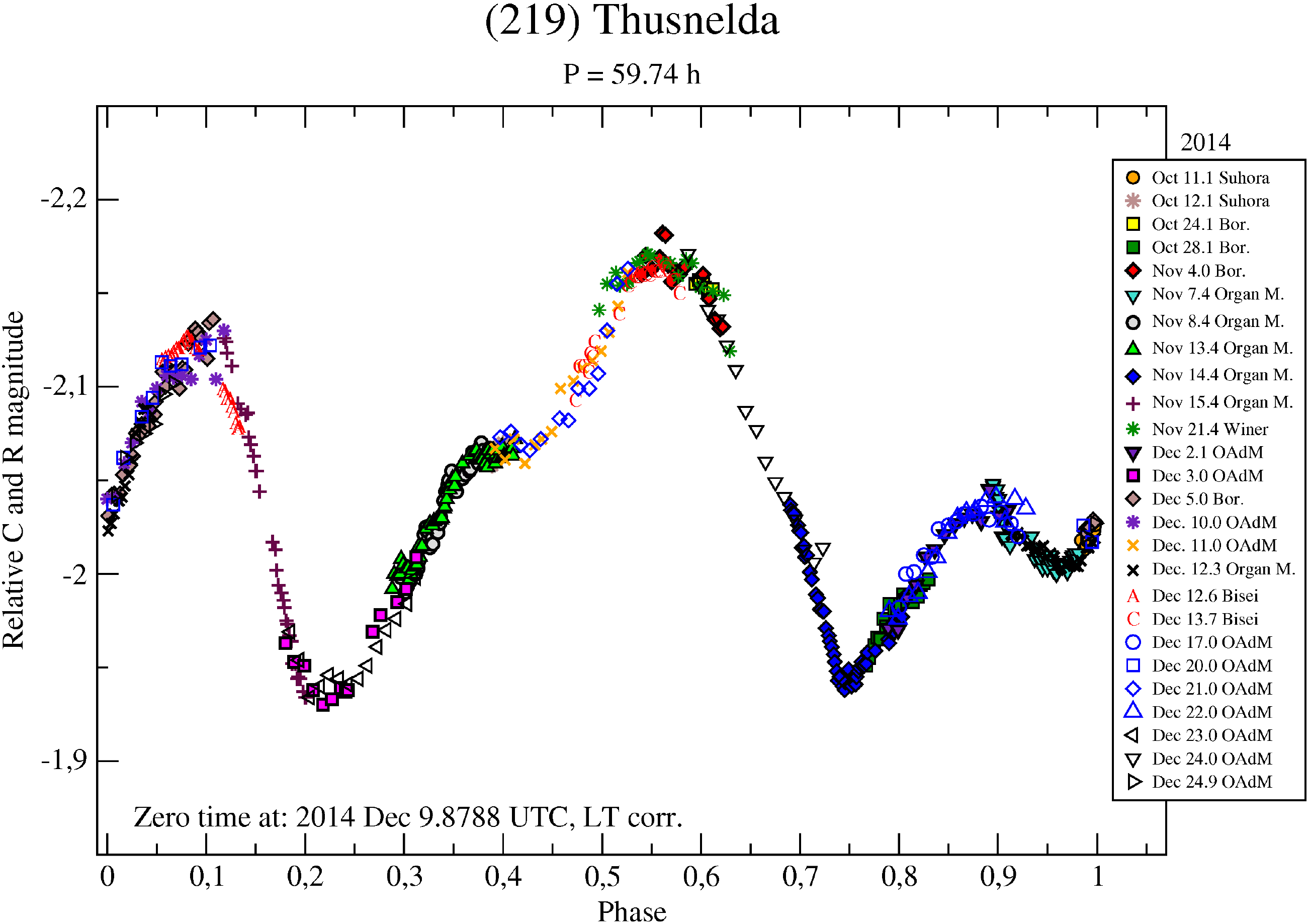}
\caption{Composite lightcurve of (219) Thusnelda in the year 2014}
\label{Thusnelda}
\end{figure}

 We observed Thusnelda in the year 2013 apparition, but the results were inconclusive in terms of the period, because of 
 too short lightcurve fragments. However, in 2014 we obtained a wealth of data, which clearly indicate 
 the period of 59.80 hours, which is close to double the value previously accepted (see figure \ref{Thusnelda}). 
 The amplitude is 0.24 $\pm$ 0.01 mag. The composite lightcurve shows interesting features with a shelf 
 before one of two maxima, and a bump in one of the minima. 
 We covered almost each phase multiple times to confirm the new period, but it took no more than 
 ten longer fragments to determine its value. Data from previous apparitions fit the new period well. 
 Almost 60-hour period, although very long, turned out not that difficult 
 to be found and fully covered, even though only relative lightcurves were used. This fact 
 is encouraging for further works on long-period asteroids. When in the low-amplitude range, they tend to 
 display irregular lightcurve features that help to resolve the period.

\subsection{(227) Philosophia}
\label{(227)Philosophia}
 
 Philosophia was observed multiple times by various authors, but the correct value of its rotation period 
 is still ambiguous. 
\citet{Bembrick_etal_2006}, 
 basing on observations from 2004 and 2005 determined period of 18.048 hours, 
 \citet{Behrend_etal_2014} 
in 2006 apparition, found the period of 26.138 hours. 
\citet{Alkema_2013a} 
 observed 
 Philosophia in 2012 and found the period of 17.181 hours. Lastly, 
\citet{Pilcher_Alkema_2014a,Pilcher_Alkema_2014b} 
 observed this asteroid in 2013/2014 apparition and published the period of 52.98 hours, but later 
 revised it to an ambiguity between 26.476 hours and 52.955 hours.
 At the beginning of our project the value accepted in the LCDB for 
 the period of Philosophia was 18.048 hours with the code "2"; now it is 52.98 hours with the code "2" and 
 a flag "A" (ambiguous).
 The amplitudes of the previously recorded lightcurves ranged from 0.06 to 0.20 mag, 
 with one determination at 0.35 mag by 
\citet{Ditteon_Hawkins_2007} 
 in 2006, which is probably erroneous. 
 The lightcurves of low amplitude were very irregular, but those of larger 
 ones were more symmetric. 
 
 It is worth noting, that in 2006 a seemingly bimodal 26.138-hour period lightcurve 
 with two clear minima was recorded 
\citep{Behrend_etal_2014}. 
 However the data for the two minima are spaced 
 by tens of days, thus a large number of rotations (39.5), and actually can cover the same minimum, 
 if the period had a slightly different value (26.46 hours proposed in this work), because then
 these minima would be spaced by exactly 39.0 cycles.
  
\begin{figure}[h]
\vspace{0.8cm}
\includegraphics[width=0.48\textwidth]{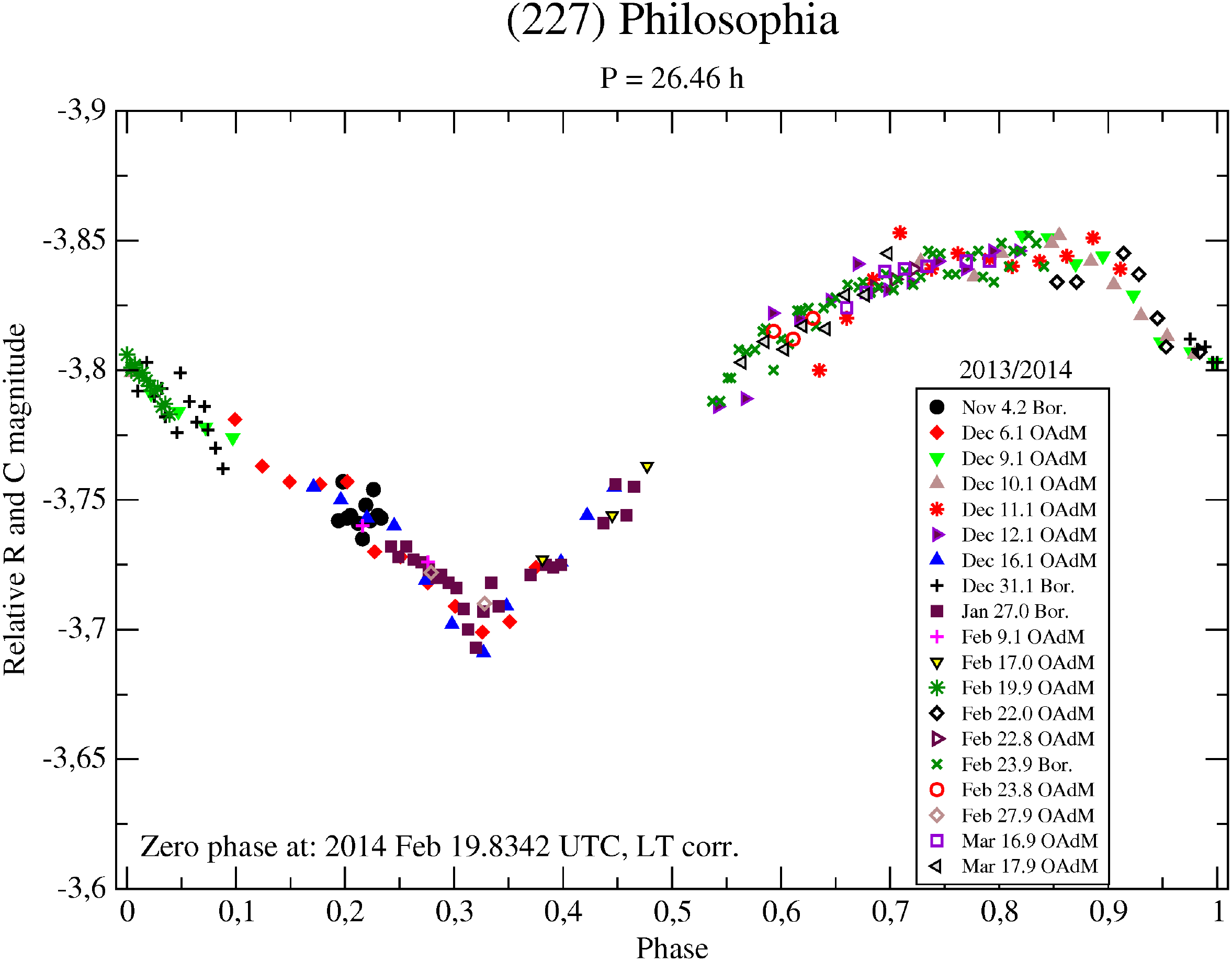}
\caption{Composite lightcurve of (227) Philosophia in the years 2013/2014}
\label{Philosophia}
\end{figure}

 In this work we observed Philosophia independently from November 2013 till March 2014 mostly using the 80-cm TJO robotic 
 telescope in the Montsec Observatory (OAdM). Basing on 19 nigths of data we found the 
 most probable period of 26.46 hours in a 0.15 mag ($\pm$ 0.02 mag) amplitude lightcurve of a monomodal character 
 (fig. \ref{Philosophia}) confirming the values published by 
\citet{Behrend_etal_2014} 
and 
\citet{Pilcher_Alkema_2014b}. 
 Our lightcurve contained small irregularities, that would repeat themselves in the second half 
 of the lightcurve if the period was forced to the double value 52.92 hours. 
 Our data from 2013/2014 apparition folded with the data by 
\citet{Pilcher_Alkema_2014a} 
from similar  time confirm the fact that bimodal, 53-hours composite lightcurve would be conspiciously symmetric 
 over half rotation, implying the same shape irregularities at both sides of the body. 
 We consider the 26 hour period the most probable also because it probably implies a switch from bimodal to monomodal 
 lightcurve character between the years 2006 and 2013/2014, a behaviour seen before in many asteroids with 
 strongly inclined spin axes. Such switch is caused by changing viewing geometry. When the asteroid 
 is viewed near an equatorial aspect it usually displays roughly regular, bimodal lightcurve,
 while viewed nearly pole-on, the asteroid displays lightcurve that is almost flat or changes its character
 to monomodal or more complicated, but always of a small amplitude. In case of Philosophia, also a change in 
 the angular distance from the ecliptic plane could play a role in the changes of the lightcurve, being 
 maximal in the year 2006 and minimal in 2013/2014.
  The 26-hour period was already shown to fit also the data from 2012 
\citep{Pilcher_Alkema_2014b}. 

\begin{figure}[h]
\vspace{0.8cm}
\includegraphics[width=0.48\textwidth]{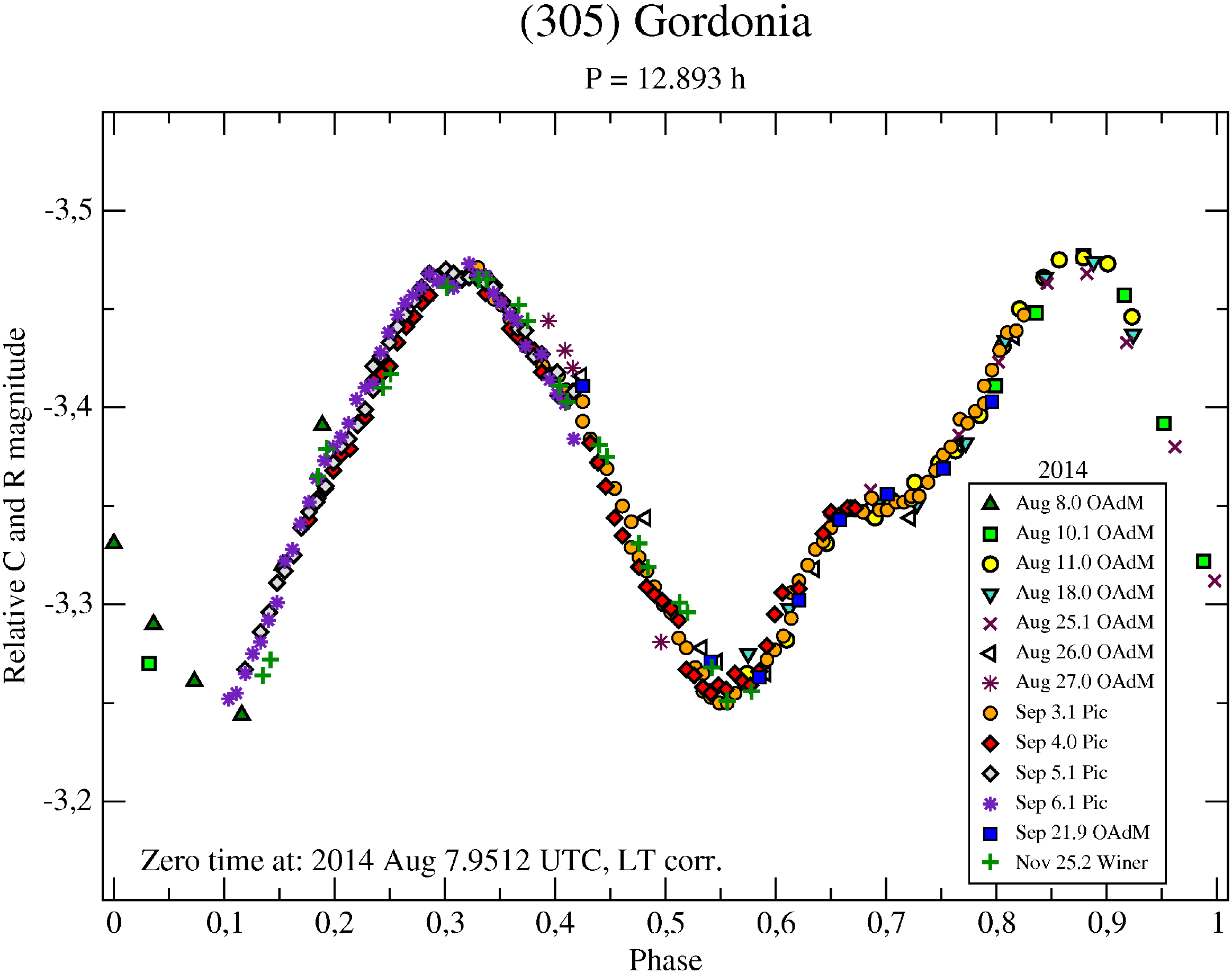}
\caption{Composite lightcurve of (305) Gordonia in the year 2014}
\label{Gordonia}
\end{figure}
 
\subsection{(305) Gordonia}

 Gordonia was observed by 
\citet{Lazar_etal_2001} 
 in the year 2000, 
\citet{Menke_etal_2008} 
in 2005, 
 and \citet{Behrend_etal_2014} 
 in 2008, and as a result three values were published for the rotation period: 16.2, 12.89, 
 and over 10 hours, respectively. In 2005, when the only full lightcurve was obtained, 
 Gordonia showed a bimodal lightcurve with wide maxima and some visible phase angle effects 
 or some instrumental artifacts. Registered amplitudes ranged from around 0.1 to 0.17 mag, 
 though it seems that the higher value should be 0.27 mag, basing on data from 2005 
\citep{Menke_etal_2008}. 
 
 We observed Gordonia in its apparition in summer 2014 and quickly realised that its period 
 must be close to 12.89 hours found by 
\citet{Menke_etal_2008}. 
The large lightcurve fragments that we obtained 
 phased with this period showed a clear, bimodal lightcurve with a small ''shelf'' before 
 one of the maxima, and an amplitude of 0.23 $\pm$ 0.02 mag (fig. \ref{Gordonia}). The final period 
 after 3.5 months of observation was determined to be 12.893 hours. The 16.2 and 10-hours 
 periods are ruled out on the basis of new data.

\subsection{(329) Svea}

Previous lightcurves of (329) Svea were published by 
\citet{Weidenschilling_etal_1990}, 
\citet{Pray_2006}, 
\citet{Behrend_etal_2014}, 
and \citet{Menke_etal_2008}. 
 The first two authors give the rotation periods of 15.0 and 15.201 
hours and the remaining two 22.770, and 22.6 hours respectively. 
The lightcurves of 15-hour periods had both a bimodal character. The lightcurves of 22-hour period
gave, trimodal lightcurve in the year 2005 
\citep{Menke_etal_2008}, 
 but bimodal one in 2006 
\citep{Behrend_etal_2014}. 

The highest quality code ''3'' in LCDB for the period determination was given to 15.201 hours determined 
by \citet{Pray_2006}, 
and that value was in the LCDB summary line in late 2012 when our target list was created 
(LCDB; \citet{Warner_etal_2009} 
). 
Later, this value was changed to 22.77 hours determined by 
\citet{Behrend_etal_2014}, 
with the quality code ''2+''.
It is worth to note that longer, 22-hour periods, were determined on richer datasets than shorter, 
15-hour ones. However, some of the previously gathered data, namely those published by 
\citet{Menke_etal_2008}, 
and \citet{Pray_2006}, 
are rather noisy or consist of short fragments, so that many different periods are possible.
\begin{figure}[h]
\vspace{0.8cm}
\includegraphics[width=0.48\textwidth]{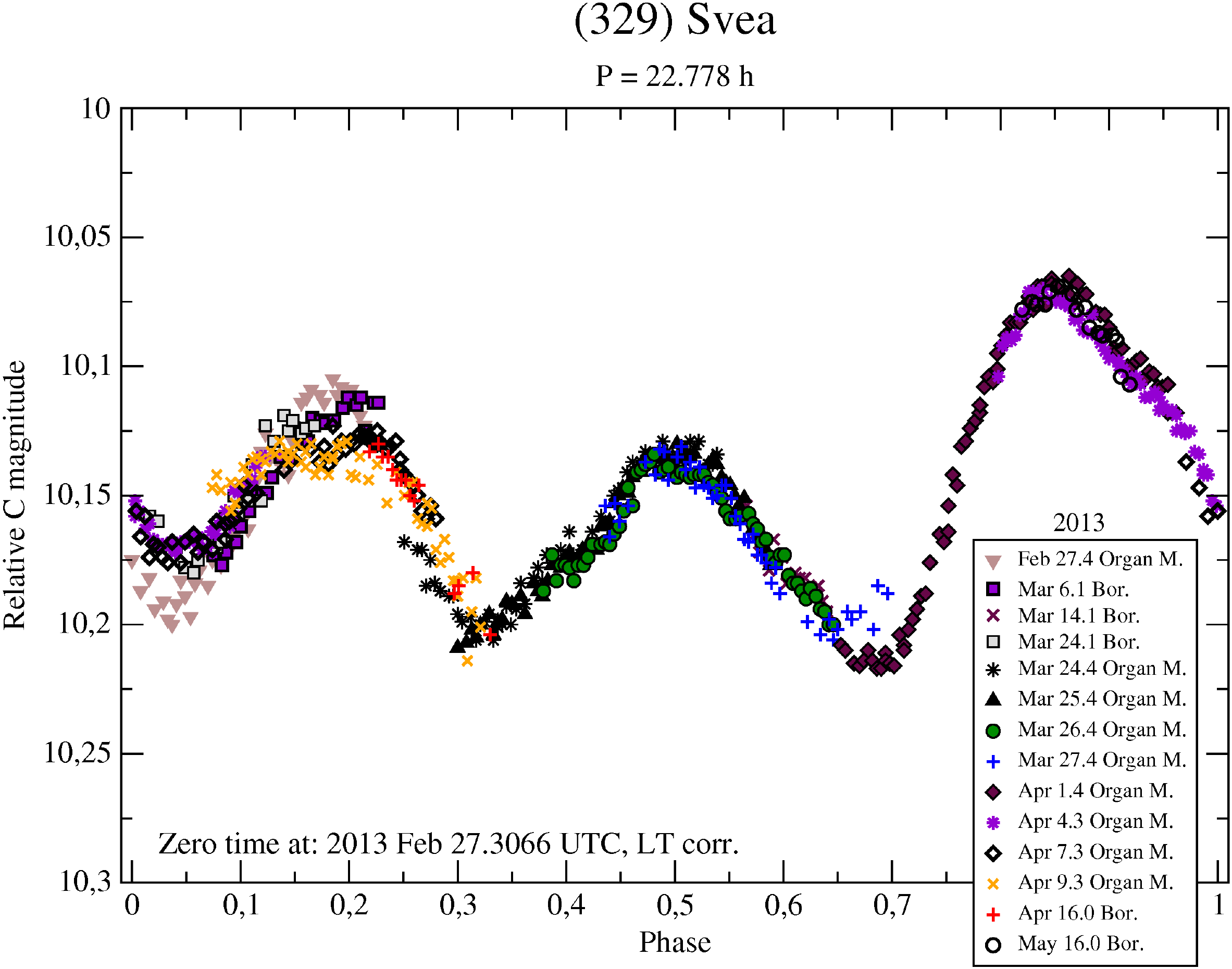}
\caption{Composite lightcurve of (329) Svea in the year 2013}
\label{Svea}
\end{figure}

Our data from the apparition in March 2013 seemed to confirm 15 hours period at first. Hovewer, 
in early April 2013 a maximum of much higher amplitude than previous two was observed. The composite lightcurve 
now could only be created with 3/2 of the previous period (namely with 22.778 hours), producing three 
different maxima per full cycle, slightly changing their shape with phase angle (fig. \ref{Svea}). 
The bimodal lightcurve of 15 hours is not possible now, as can be inferred from the periodogram 
(fig. \ref{329periods}). The new, longer period seems to depend on a single night of 1 April, but even without this night 
a much better fit is obtained for 22 hours, than for 15 hours period.

This way we confirmed the values given by 
\citet{Behrend_etal_2014} 
and 
\citet{Menke_etal_2008}. 
The phase angle effects are especially visible in the shape and height of the first maximum 
in fig. \ref{Svea}. First, in February 2013, two months before the opposition date, this maximum was 
higher and peaked, then in March a notch on the top started to mark itself to become more evident in April,  
when the amplitude was smallest, as expected near the opposition. Such effects are not so profound in the 
remaining part of the lightcurve, because those fragments were obtained with much less time span 
between separate fragments.
The amplitude in 2013 was 0.16 $\pm$ 0.02 mag, and the composite lightcurve was created on 14 fragments 
most of which covered 0.2 - 0.3 of the full cycyle.
The behaviour of this target apparently changes from simple bi-modal, to tri-modal depending on 
an aspect, indicating an interesting shape and an inclined spin axis. 

\subsection{(439) Ohio}

Observed previously in the year 1984 by 
\citet{Lagerkvist_etal_1987}, 
 asteroid Ohio displayed 
one wide maximum and a flat minimum in a 19.2 hour lightcurve, with an amplitude of 0.24 mag. 
There are also data gathered by group led by 
\citet{Behrend_etal_2014} 
in 2004 and 2007
but these lightcurves were highly incomplete. A possibility of a bimodal lightcurve behaviour was mentioned by the first authors, 
if the period was twice as long, namely 38.4 hours.

\begin{figure}[h]
\vspace{0.8cm}
\includegraphics[width=0.48\textwidth]{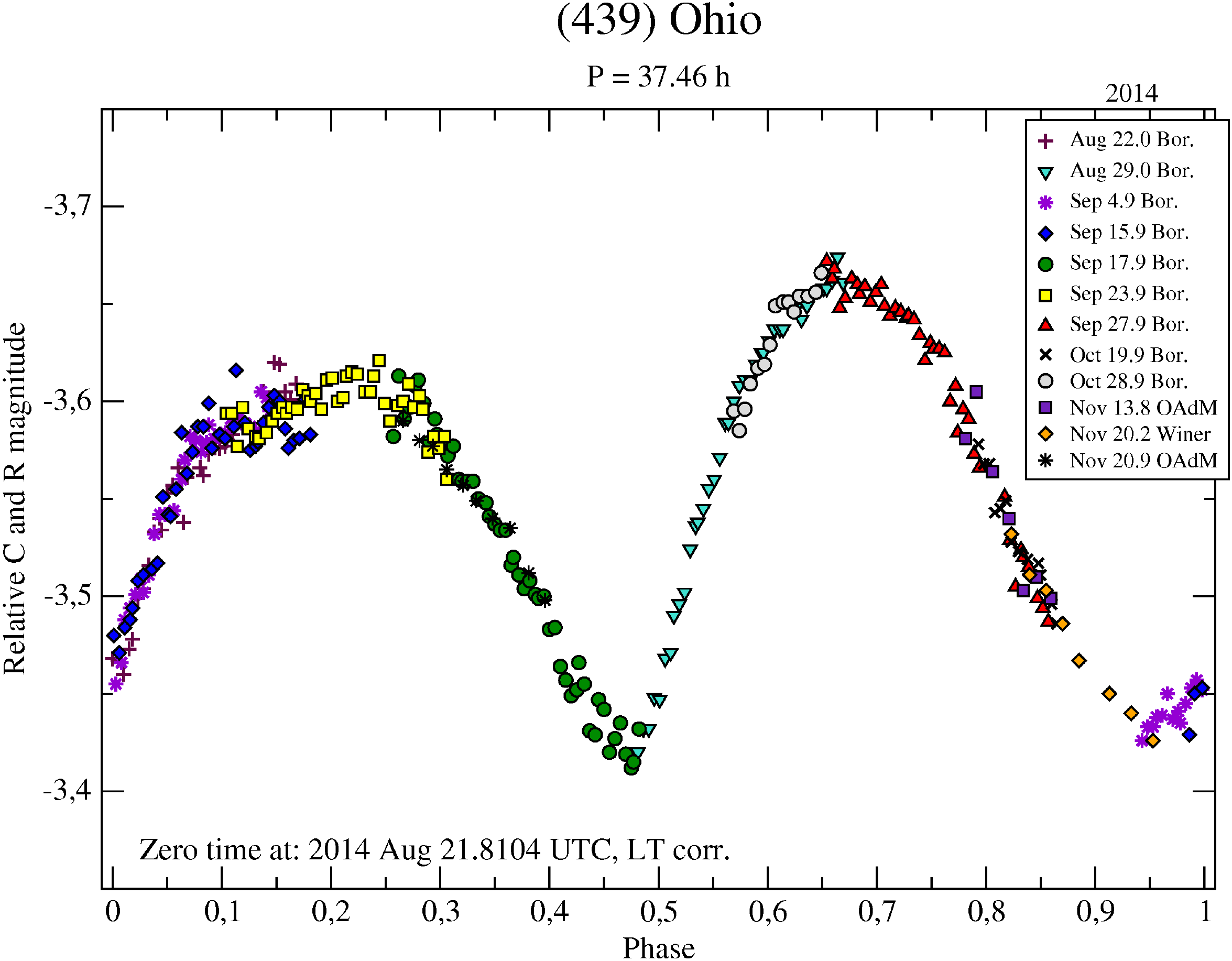}
\caption{Composite lightcurve of (439) Ohio in the year 2014}
\label{Ohio}
\end{figure}

We confirmed that supposition, observing this target from August till November 2014. We found the best fit period an hour shorter 
than suggested by 
\citet{Lagerkvist_etal_1987}, 
at 37.46 hours, in a bimodal, slightly asymmetric lightcurve 
of 0.23 $\pm$ 0.02 mag amplitude (fig. \ref{Ohio}). The short period, 19.2 hours, can be safely rejected basing on new data.

\subsection{(666) Desdemona}

Desdemona was previously observed in three apparitions: in the year 2000 
\citep{Stephens_2001}, 
and in 2004 and 2006/2006 
\citep{Behrend_etal_2014}. 
Here too, only the first publication contains full phase coverage, the period was determined at 15.45 hours, and an amplitude 
at 0.11 mag. Data from next two appartions suggested 9.2-hours period and amplitudes ranging from 0.07 to 0.16 mag.

\begin{figure}[h]
\vspace{0.8cm}
\includegraphics[width=0.48\textwidth]{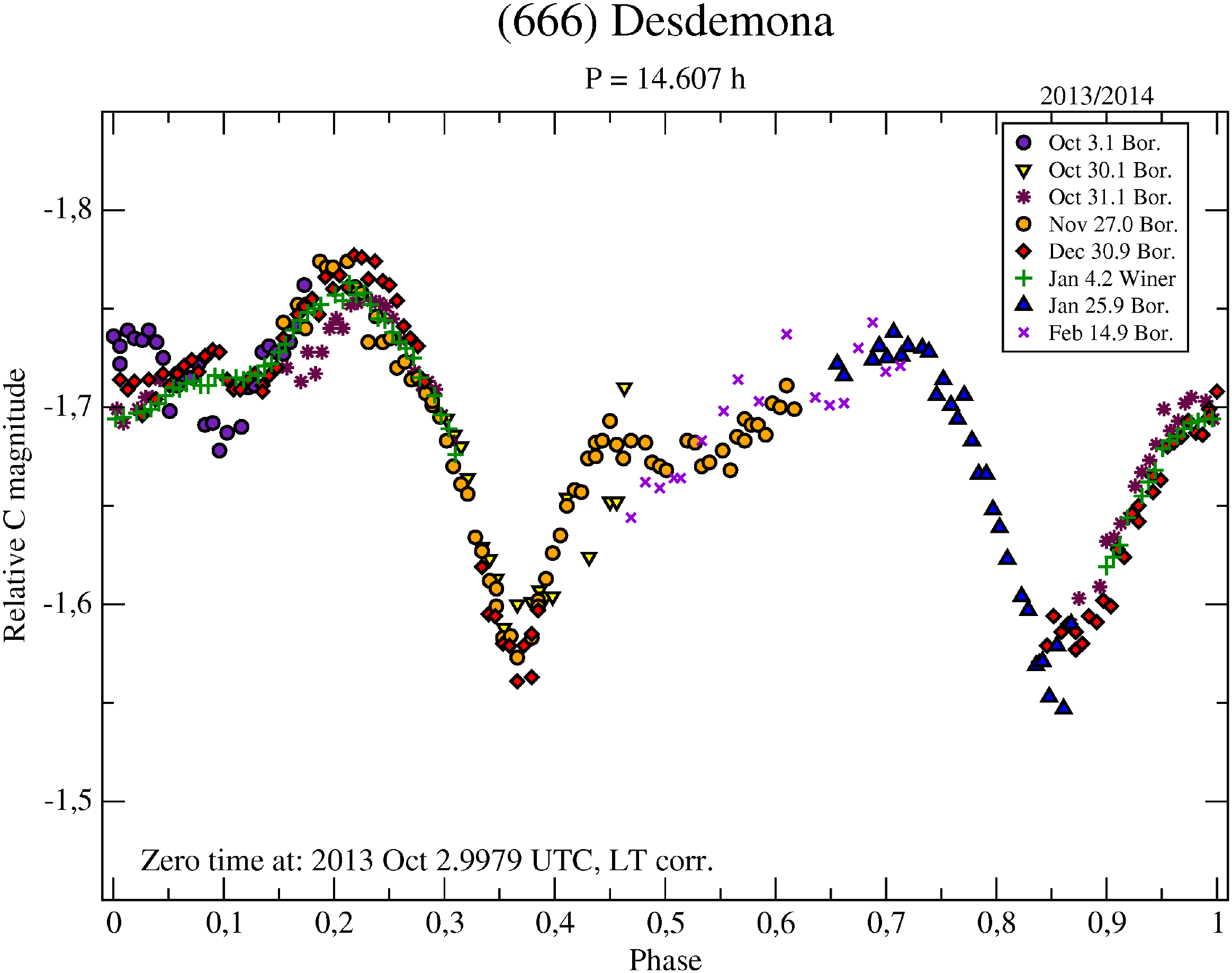}
\caption{Composite lightcurve of (666) Desdemona in the years 2013/2014}
\label{Desdemona}
\end{figure}

We observed Desdemona from October 2013 till February 2014, and had problems finding its period at first. Prevously published 
values did not fit the new data, and the relatively good fit could only be obtained on the basis of 14.607 hours, if some discrepancies 
could be assigned to phase angle effect (see fig. \ref{Desdemona}). This is possible, since most overlapping fragments differ in time 
by a month or more. The composite lightcurve of this target contains two narrow minima and very wide, wavy maxima, and has an 
amplitude of 0.22 $\pm$ 0.02 mag.


\subsection{(806) Gyldenia}
Previously gathered lightcurves of this target come from the years 2005, 2007 
\citep{Behrend_etal_2014}, 
and 2013 
\citet{Alkema_2013b}. 
Data from both 2005, and 2007 suggest a period of 14.45 hours, but have incomplete coverage, and probably some 
observing artifacts in a form of sudden dimming at the end of the observing sessions. Apart from it, the amplitudes 
were 0.27 and 0.10 mag, respectively. Data published by 
\citet{Alkema_2013b} 
 fit a different period: 16.846 hours, 
and have a multiple phase coverage in a lightcurve of 0.14 mag amplitude. The new data were shown to misfit the period 
of 14.52 hours. In the summary line in LCDB database the old value of period was changed to the new one, with a code 3-.

\begin{figure}[h]
\vspace{0.8cm}
\includegraphics[width=0.48\textwidth]{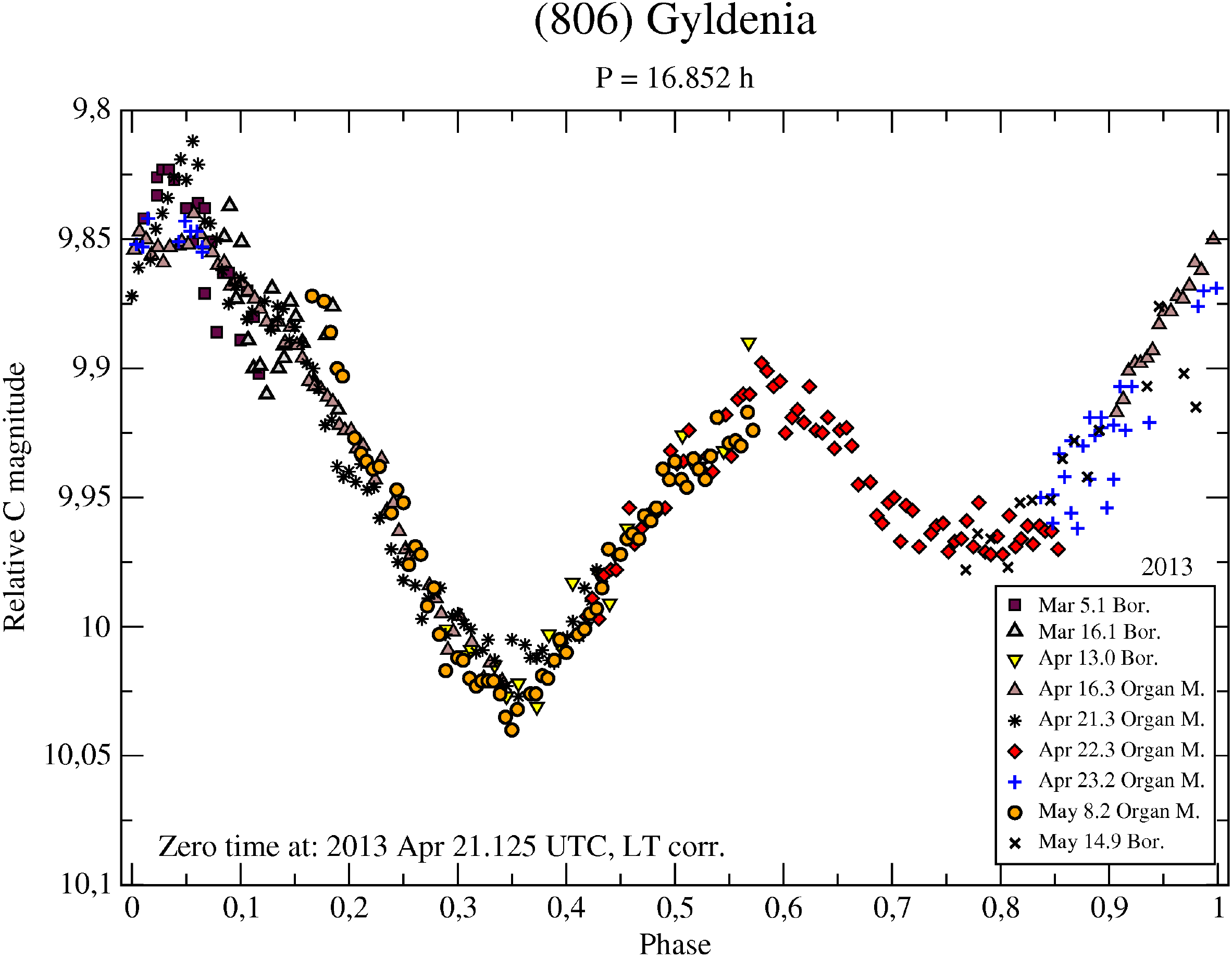}
\caption{Composite lightcurve of (806) Gyldenia in the year 2013}
\label{Gyldenia}
\end{figure}

We observed Gyldenia independently in 2013, and also found the old period value to be wrong. We found 16.856 hours to fit best, 
which is a similar value to that published by 
\citet{Alkema_2013b}. 
However, the amplitude of our lightcurve is substantially larger: 
0.20 $\pm$ 0.02 mag (fig. \ref{Gyldenia}), though it was determined in the same apparition. There are two fragments in our composite 
lightcurve that stretch from peak to peak, and they were obtained under small phase angle (2$^\circ$ and 3$^\circ$), 
so their amplitudes were not enlarged by phase angle effects. 
We suppose that the lower amplitude determined by 
\citet{Alkema_2013b} 
 was caused by short fragments that could be arbitrarily 
shifted in vertical scale what suppressed the amplitude. The overall look of our composite lightcurve is highly asymmetric, 
with maxima and minima at notably different levels.

\section{Conclusions and future work}

We came across substantial selection effects among bright main-belt asteroids and started a campaign to reduce them. 
Our survey resulted in finding new values of the rotation period for a quarter of the studied sample of long-period and 
low-amplitude asteroids. 
Majority of the revised periods occured to be longer than it seemed previously. The diameters of our current targets are 
mostly large, around a few tens of kilometers. Since the timescale of YORP effectiveness is proportional to the radius squared 
\citep{Rubincam_2000}, 
in the size range studied here this effect is probably negligible. Also, the damping timescales for tumbling 
asteroids in this size range are a fraction of the age of the Solar System 
\citep{Pravec_etal_2005}. 
So the low rotation frequencies found here are most probably primordial, or can in part be an outcome of major collisions 
in the epoch when the Main Belt was much more populated. Basing on these facts we expect mostly prograde rotations of these 
asteroids, as an outcome of growth from small pebles embedded in the protoplanetary gas disc 
\citep{Johansen_Lacerda_2010}, 
and no substantial trends for spin axis positions.
Low frequency also implies that the internal structure can be more loosely bound than if the frequency was higher. 
Consequently, the expected bulk densities in this class of objects can be lower, implying larger macroporosity. 
The shapes of these objects have low probability of being rotationally reshaped or disrupted. Their asymmetric lightcurves 
imply irregular shapes. 

Our findings strenghten the need to study such targets, even though they are observationally demanding. We are planning 
to observe these, and also fainter (smaller) targets in future apparitions, until the data suffice for obtaining 
unique spin and shape models. 
The improved statistics of periods, spins, and shapes will allow making more robust conclusions on the present state 
and previous history of the Solar System with both violent collisions and subtle thermal forces influencing those minor bodies. 

\section*{Acknowledgements}

\begin{scriptsize}
       This work was partialy supported by grant no. 2014/13/D/ST9/01818, DO was supported by grant no.  
       2012/04/S/ST9/00022, KK by grant no. UMO-2011/01/D/ST9/00427, and MP by grant no. 2014/13/B/ST9/00902 
       all from the National Science Centre, Poland. 
       The work of TSR was carried out through the Gaia Research for European Astronomy Training (GREAT-ITN) 
       network. He received funding from the European Union Seventh Framework Programme (FP7/2007-2013)
       under grant agreement no.264895.
    
    The Joan Oró Telescope (TJO) of the Montsec Astronomical Observatory (OAdM) 
    is owned by the Catalan Government and operated by the Institute for Space Studies of Catalonia (IEEC).
    
    Data from Pic du Midi Observatory have been obtained with the 0.6 m telescope, a facility operated by observatoire 
    Midi-Pyr{\'e}n{\'e}es and Association T60, an amateur association.
    We thank N. Takahashi, and the staff members of Bisei Spaceguard Center for their support. 
    We also acknowledge the Japan Space Forum.
\end{scriptsize}


{}

\appendix{}

\section{Appendix}

Period spectra for targets with revised periods.

\begin{figure}[h]
\includegraphics[width=0.45\textwidth]{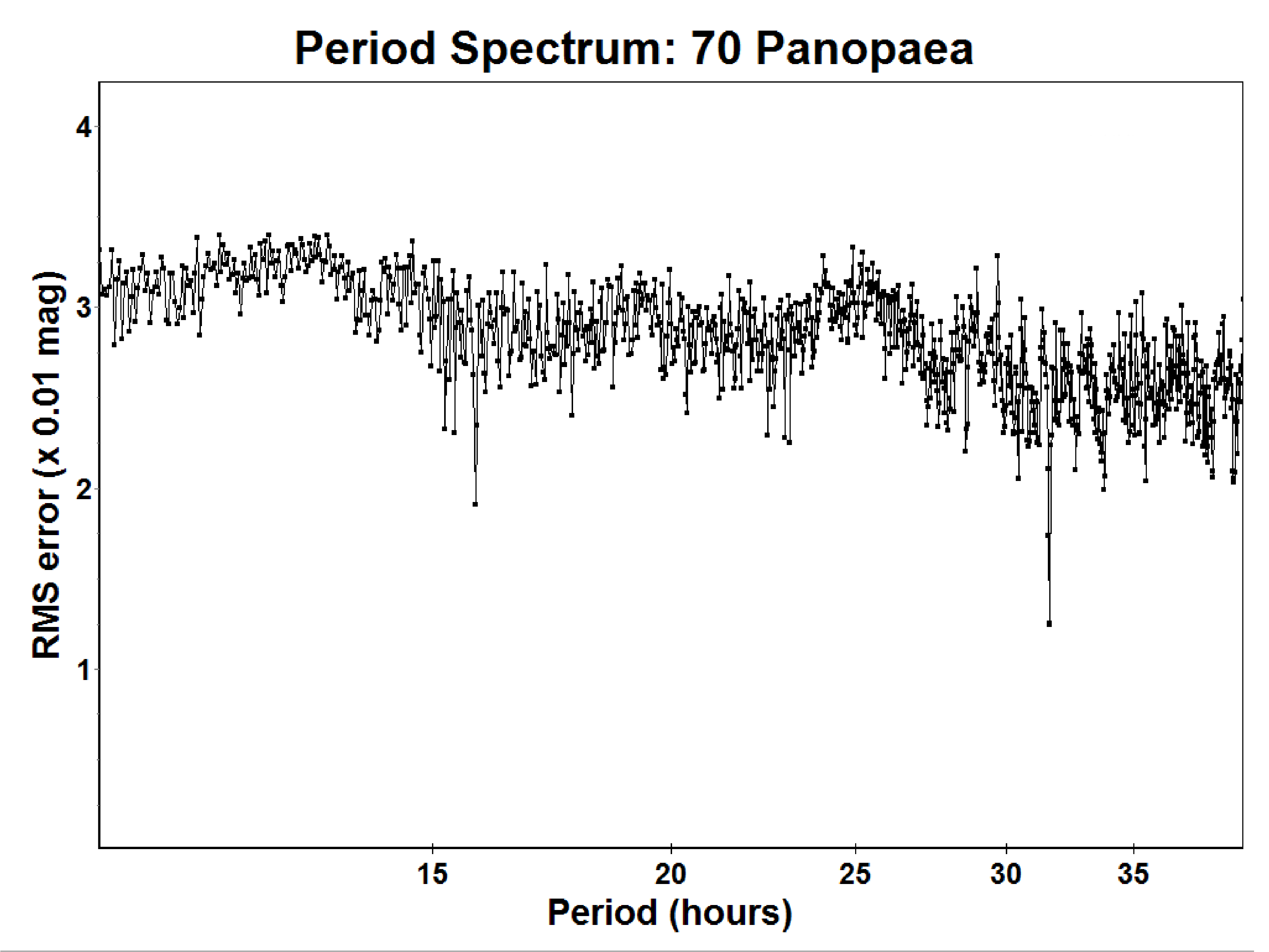}
\caption{Period spectrum for (70) Panopaea based on data from the year 2014. Lowest rms for P=31.619 hours.}
\label{70periods}
\end{figure}

\begin{figure}[h]
\includegraphics[width=0.45\textwidth]{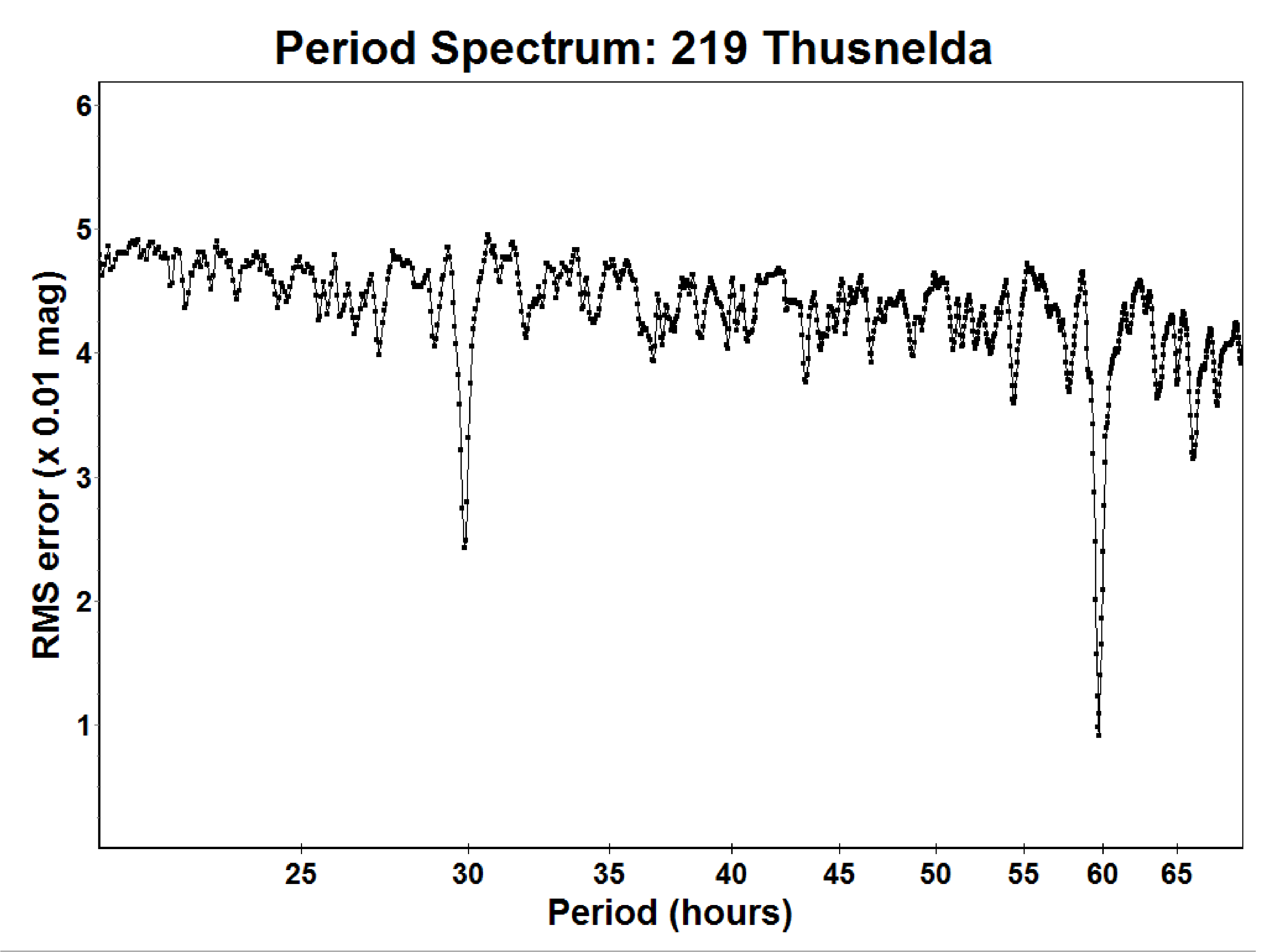}
\caption{Period spectrum for (219) Thusnelda based on data from the year 2014. Lowest rms for P=59.74 hours.}
\label{219periods}
\end{figure}

\begin{figure}[h]
\includegraphics[width=0.45\textwidth]{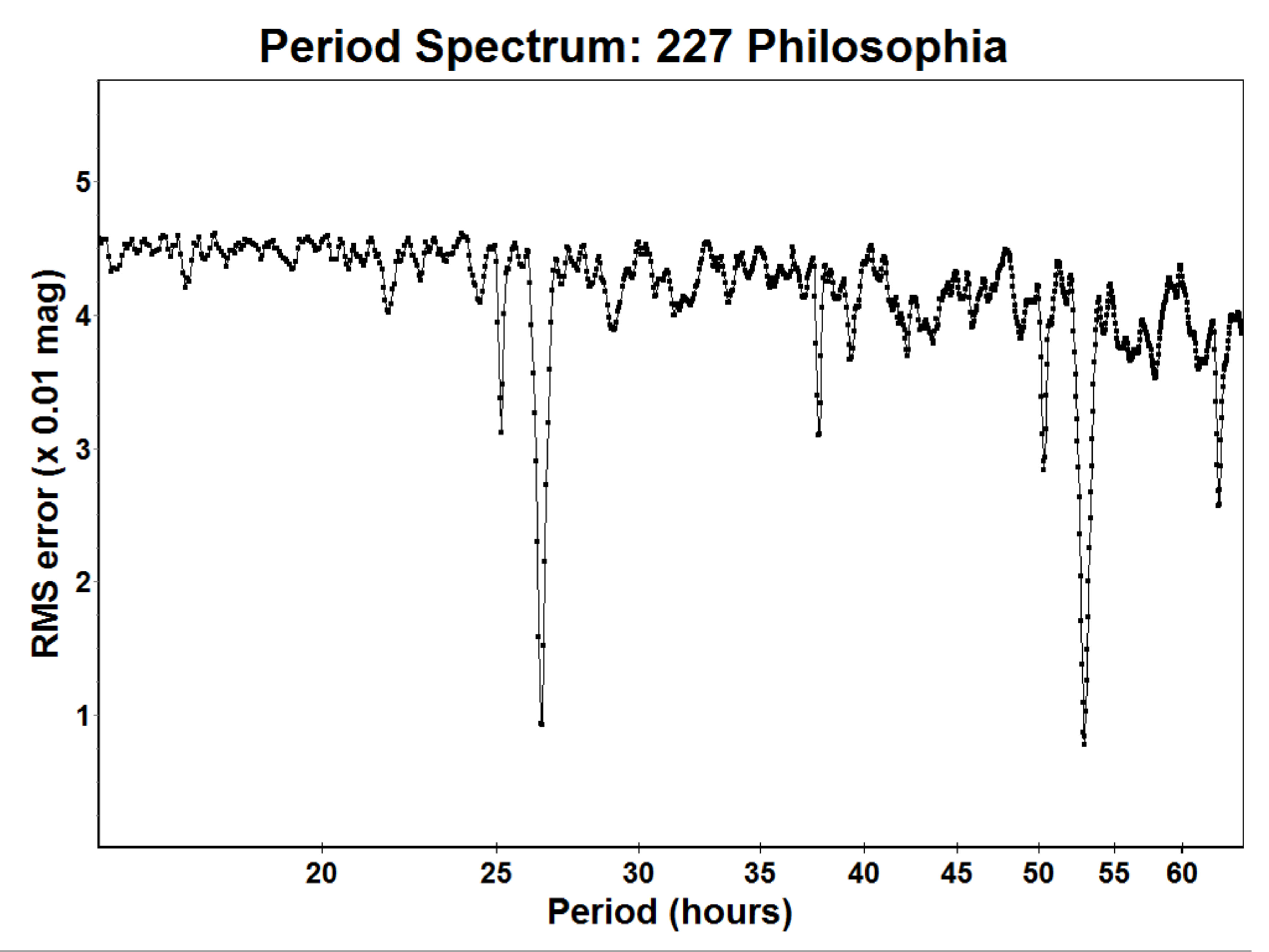}
\caption{Period spectrum for (227) Philosphia based on data from the years 2013/2014. Accepted period P=26.46 hours.}
\label{227periods}
\end{figure}

\begin{figure}[h]
\includegraphics[width=0.45\textwidth]{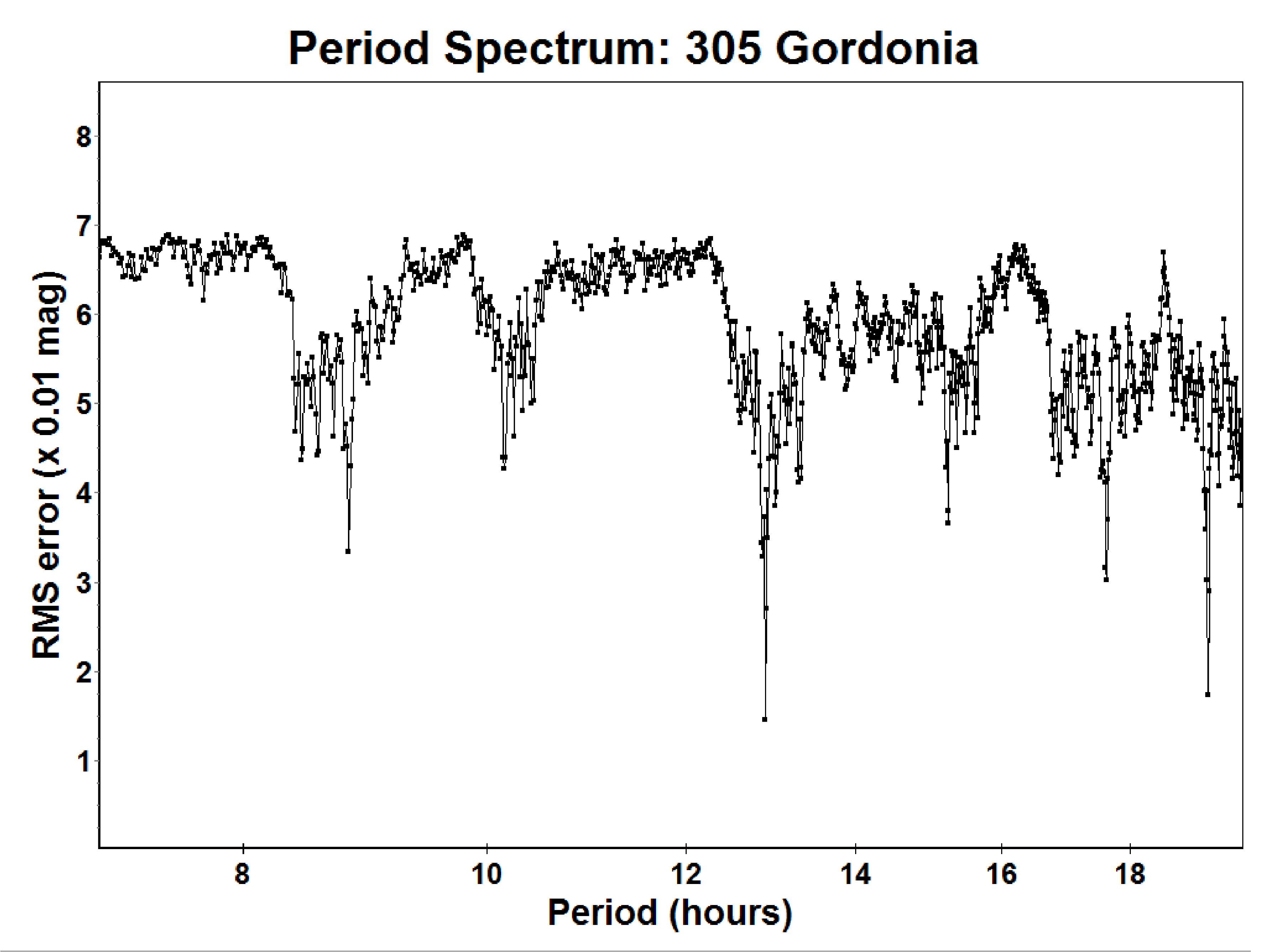}
\caption{Period spectrum for (305) Gordonia based on data from the year 2014. Lowest rms for P=12.893 hours.}
\label{305periods}
\end{figure}

\begin{figure}[h]
\includegraphics[width=0.45\textwidth]{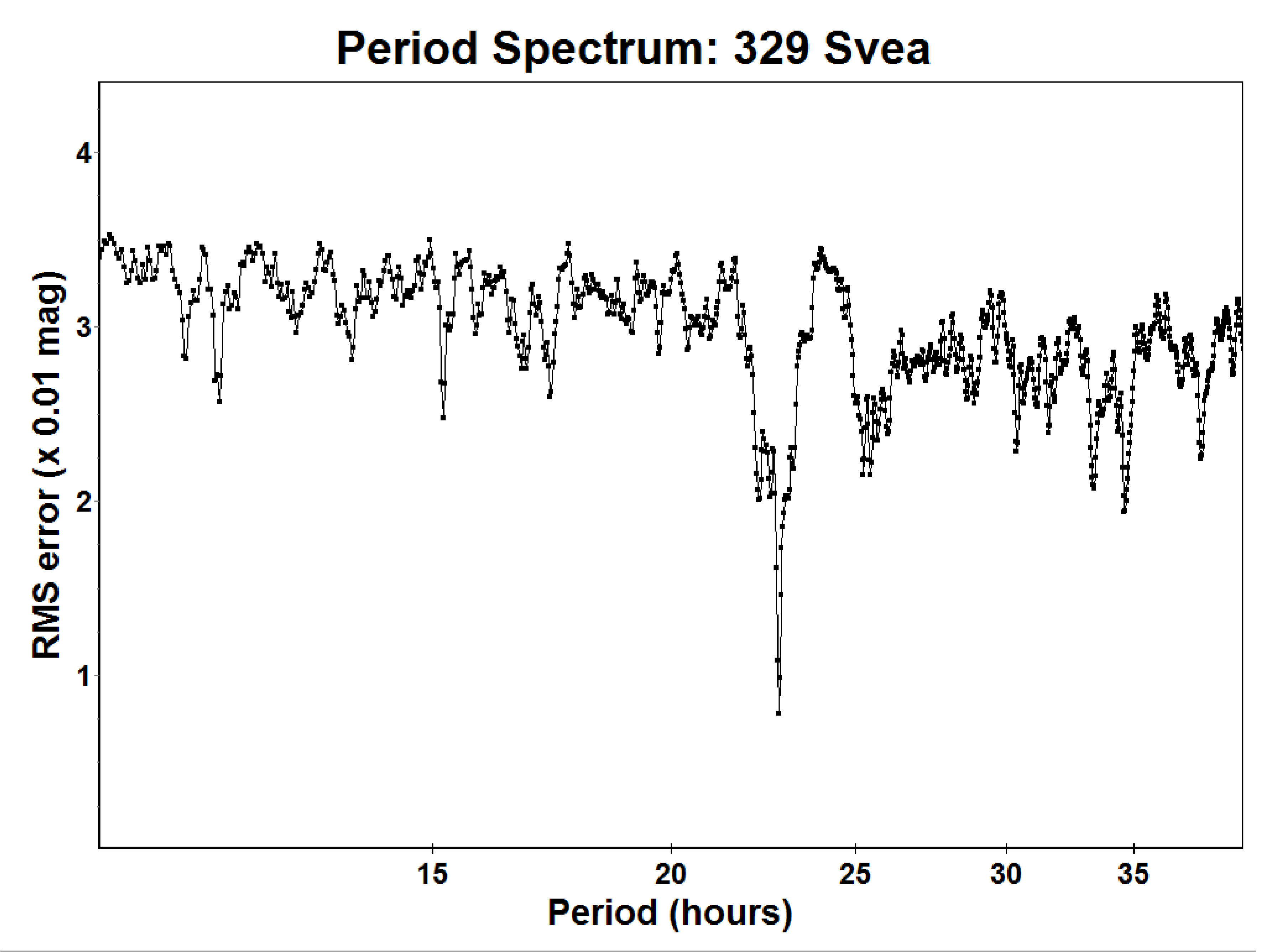}
\caption{Period spectrum for (329) Svea based on data from the year 2013. Lowest rms for P=22.778 hours.}
\label{329periods}
\end{figure}

\begin{figure}[h]
\includegraphics[width=0.45\textwidth]{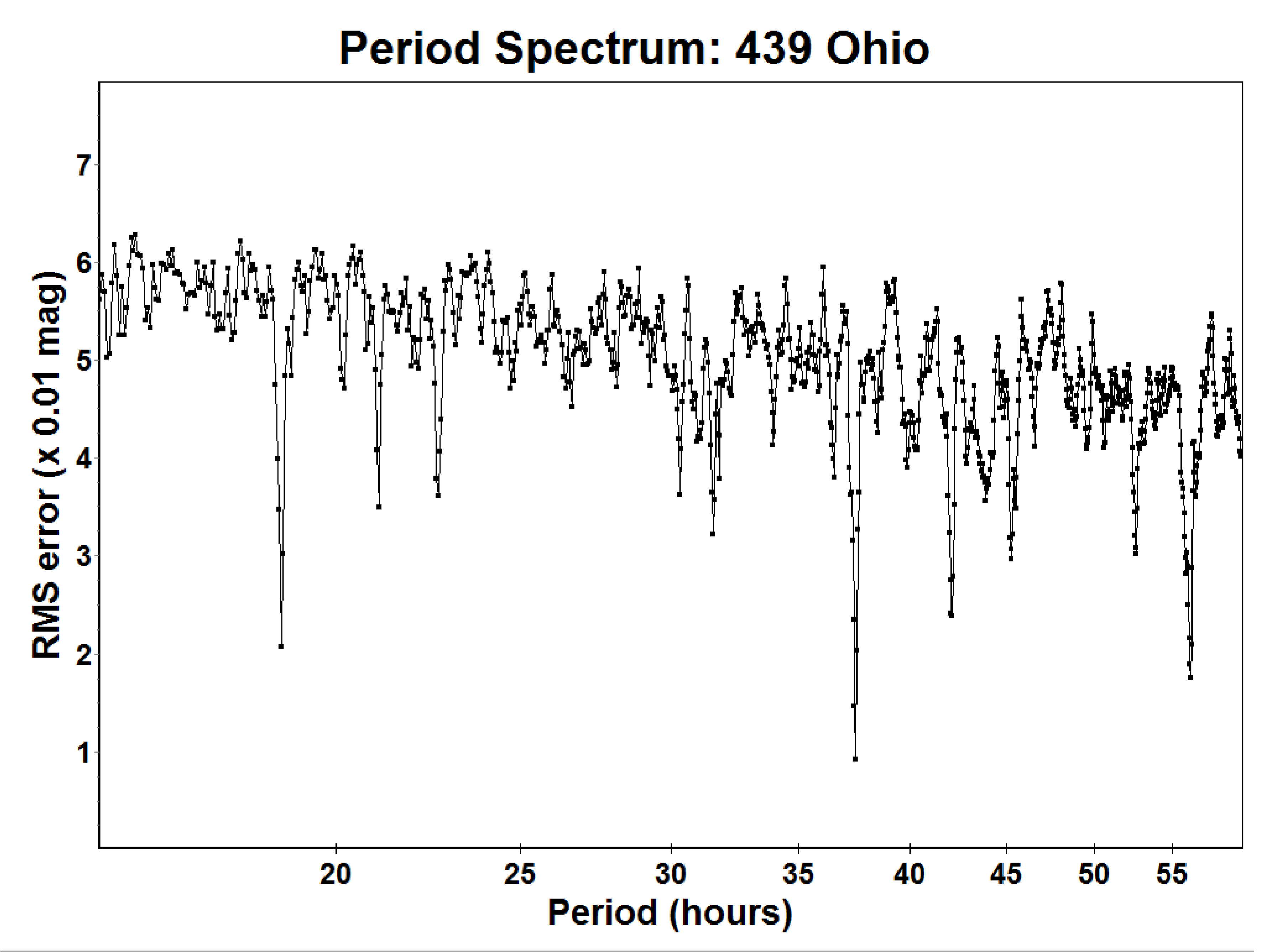}
\caption{Period spectrum for (439) Ohio based on data from the year 2014. Lowest rms for P=37.46 hours.}
\label{439periods}
\end{figure}

\begin{figure}[h]
\includegraphics[width=0.45\textwidth]{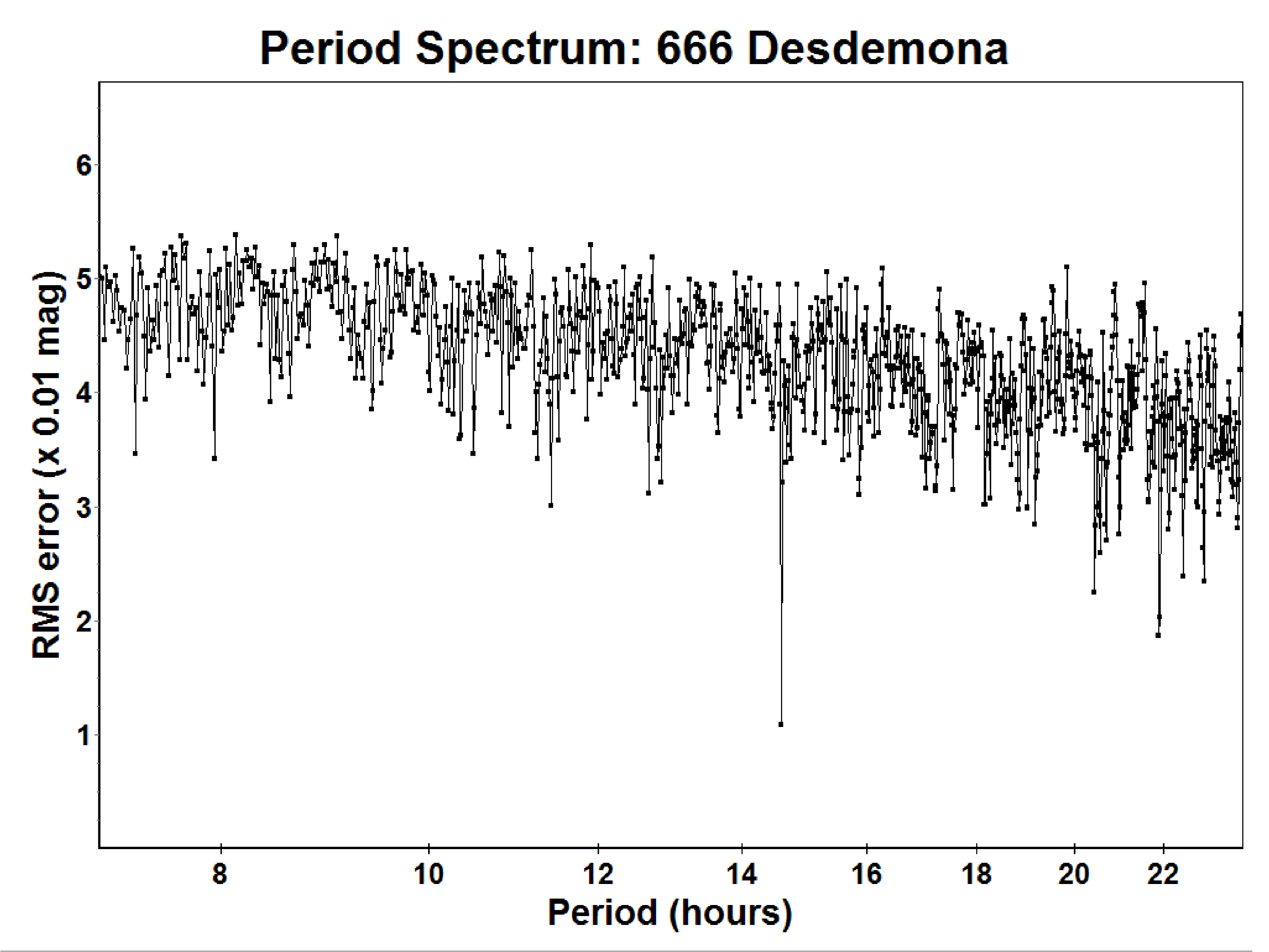}
\caption{Period spectrum for (666) Desdemona based on data from the years 2013/2014. Lowest rms for P=14.607 hours.}
\label{666periods}
\end{figure}

\begin{figure}[h]
\includegraphics[width=0.45\textwidth]{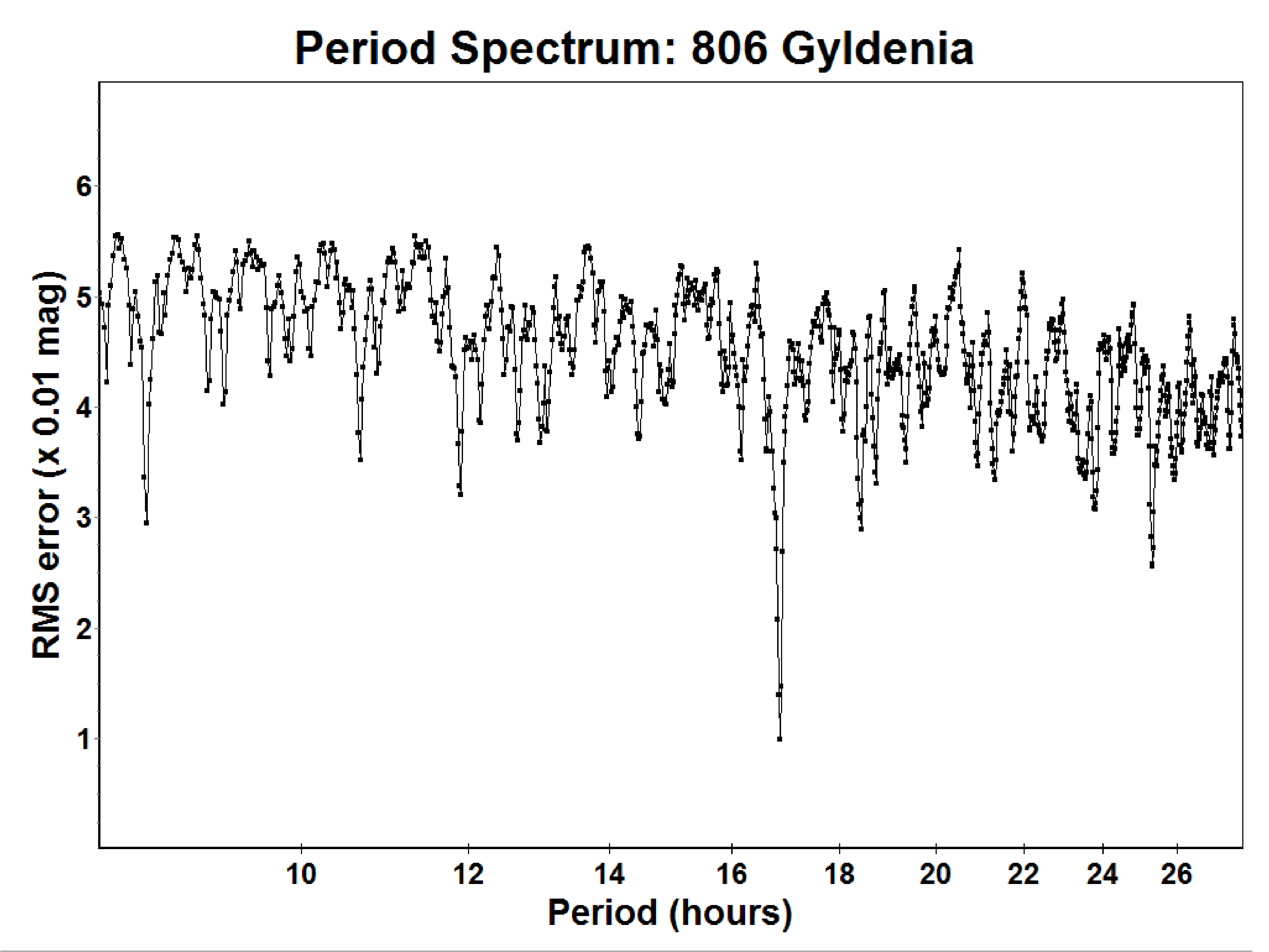}
\caption{Period spectrum for (806) Gyldenia based on data from the year 2013. Lowest rms for P=16.852 hours.}
\label{806periods}
\end{figure}

\end{document}